\documentclass[10pt,twocolumn,floatfix,aps,pra,superscriptaddress]{revtex4-2}
\usepackage{amsmath,amssymb,amsthm,mathrsfs,amsfonts,dsfont,amstext} 
\usepackage{textcomp,pbox}
\usepackage[export]{adjustbox}
\usepackage{soul} 
\usepackage{bm,xspace}
\usepackage{dcolumn,booktabs,url}
\usepackage[scaled]{helvet}
\usepackage{sansmath,gensymb}
\usepackage{tikz,graphicx,transparent,color}
\usepackage{multirow}
\usepackage[separate-uncertainty = true]{siunitx}
\usepackage{comment}
\usepackage{physics}
\usepackage{pdfpages}

\usepackage{array}
\usepackage{makecell}
\usepackage{tabularx}
\usepackage{dsfont}

\newcolumntype{L}{>{\raggedright\arraybackslash}X}

\makeatletter
\AtBeginDocument{\let\LS@rot\@undefined}
\makeatother

\newcolumntype{C}[1]{>{\middleing\let\newline\\\arraybackslash\hspace{0pt}}m{#1}}

\usepackage{natbib} 

\usepackage[colorlinks=true]{hyperref}
\usepackage{graphicx}

\hypersetup{
     colorlinks   = true,
     citecolor    = red,
     linkcolor    = blue,
     urlcolor     = red     
}

\graphicspath{{./figs/}}

\makeatletter
\def\maketitle{
\@author@finish
\title@column\titleblock@produce
\suppressfloats[t]}
\makeatother

\usepackage[resetlabels]{multibib}

\newcites{S}{References}

\usepackage{lineno}

\begin{document}

\newcommand{\TitleName}{Tailoring fusion-based photonic quantum computing schemes to quantum emitters}
\title{\TitleName}


\author{Ming Lai Chan$^{\dagger, }$}
\email{Ming-Lai.Chan@sparrowquantum.com}
\affiliation{Center for Hybrid Quantum Networks (Hy-Q), The Niels Bohr Institute, University~of~Copenhagen,  DK-2100  Copenhagen~{\O}, Denmark}
\affiliation{Sparrow Quantum, Blegdamsvej 104A, DK-2100  Copenhagen~{\O}, Denmark}

\author{Thomas J. Bell}
\thanks{These authors contributed equally to this work.}
\affiliation{Quantum Engineering Technology Labs, H. H. Wills Physics Laboratory and Department of Electrical and
Electronic Engineering, University of Bristol, BS8 1FD, United Kingdom}
\affiliation{Quantum Engineering Centre for Doctoral Training, University of Bristol, UK}

\author{Love A. Pettersson}

\affiliation{Center for Hybrid Quantum Networks (Hy-Q), The Niels Bohr Institute, University~of~Copenhagen,  DK-2100  Copenhagen~{\O}, Denmark}

\author{Susan X. Chen}
\affiliation{Quantum Engineering Centre for Doctoral Training, University of Bristol, UK}
\affiliation{NNF Quantum Computing Programme, Niels Bohr Institute, University of Copenhagen, Blegdamsvej 17, DK-2100 Copenhagen Ø, Denmark}

\author{Patrick Yard}

\affiliation{Quantum Engineering Technology Labs, H. H. Wills Physics Laboratory and Department of Electrical and
Electronic Engineering, University of Bristol, BS8 1FD, United Kingdom}

\author{Anders S. Sørensen}

\affiliation{Center for Hybrid Quantum Networks (Hy-Q), The Niels Bohr Institute, University~of~Copenhagen,  DK-2100  Copenhagen~{\O}, Denmark}

\author{Stefano Paesani}
\email{stefano.paesani@nbi.ku.dk}

\affiliation{NNF Quantum Computing Programme, Niels Bohr Institute, University of Copenhagen, Blegdamsvej 17, DK-2100 Copenhagen Ø, Denmark}
\affiliation{Center for Hybrid Quantum Networks (Hy-Q), The Niels Bohr Institute, University~of~Copenhagen,  DK-2100  Copenhagen~{\O}, Denmark}

\date{}


\begin{abstract}
Fusion-based quantum computation is a promising quantum computing model where small-sized photonic resource states are simultaneously entangled and measured by fusion gates. 
Such operations can be readily implemented with scalable photonic hardware: resource states can be deterministically generated by quantum emitters and fusions require only shallow linear-optical circuits. 
Here, we propose fusion-based architectures tailored to the capabilities and noise models in quantum emitters. 
We show that high tolerance to dominant physical error mechanisms can be achieved, with fault-tolerance thresholds of 8\% for photon loss, 4\% for photon distinguishability between emitters, and spin noise thresholds well above memory-induced errors for typical spin-photon interfaces. 
Our construction and analysis provide guidelines for the development of photonic quantum hardware targeting fault-tolerant applications with quantum emitters.
\end{abstract}

\date{\today}

\maketitle


\section{Introduction}

Photonics has emerged as a leading platform for fault-tolerant quantum computing (FTQC), due to its scalability and inherently long coherence times.
These benefits come at the cost of probabilistic entangling gates, and FTQC with photonic qubits therefore motivates a different approach than the circuit-based model favoured by other platforms. 
Measurement-based quantum computing (MBQC)~\cite{Raussendorf2001} is a more natural model, exchanging entangling gates with destructive measurements on large and highly entangled cluster states.
A variant of this approach, 
fusion-based quantum computation (FBQC), was recently proposed~\cite{bartolucci_fusion-based_2023}, in which two-qubit \textit{fusion} measurements are systematically applied to small entangled states to effectively entangle and measure the cluster state of an MBQC scheme in a single step.
The \textit{resource states} of FBQC can therefore be much smaller, reducing overheads of a photonic quantum computer, but their generation remains a critical and costly part of the architecture.
Quantum emitters, such as neutral atoms, semiconductor quantum dots or colour centers, have been shown to be promising candidates for the deterministic generation of multiphoton entangled states \textit{on-demand}~\cite{lindner_proposal_2009, Schwartz2016, Thomas2022,Cogan2023,Coste2023, Meng2024}, which could lead to lower resource costs as compared to sources based on spontaneous parametric down conversion and probabilistic entangling gates~\cite{Zhong2018}.

Photon loss remains the primary obstacle to the practical implementation of FBQC schemes, and it is therefore critical to ensure robustness of any photonic architecture to this failure mode.
Previous works using unencoded resource states and boosted physical fusions can typically tolerate loss below $1\%$~\cite{sahay_tailoring_2022,paesani_high-threshold_2022}, a daunting prospect for current state-of-the-art devices.
Code concatenation is an effective method to increase loss-tolerance, and schemes using Shor codes~\cite{bartolucci_fusion-based_2023, pankovich_high_2023, Song2024} and graph codes~\cite{Bell_2023_optimizing, Pettersson2025} have been used to achieve great improvements in the loss thresholds of FBQC.
Many of these proposals also feature adaptivity in the ordering or basis of fusion measurements~\cite{bombin_increasing_2023, Song2024, Bell_2023_optimizing}, achieving loss thresholds in excess of 10\%, but the assumed complex resource states are challenging to generate, requiring multiple quantum emitters and probabilistic gates~\cite{Hilaire2023neardeterministic}.
A recent proposal based on repeat-until-success (RUS) $CZ$ gates between quantum emitters to implement a surface code logical memory has shown a modest loss threshold of 2.75\%~\cite{deGliniasty2024spinopticalquantum}.%

\begin{figure*}[t]
    \centering
    \includegraphics[width=\textwidth]{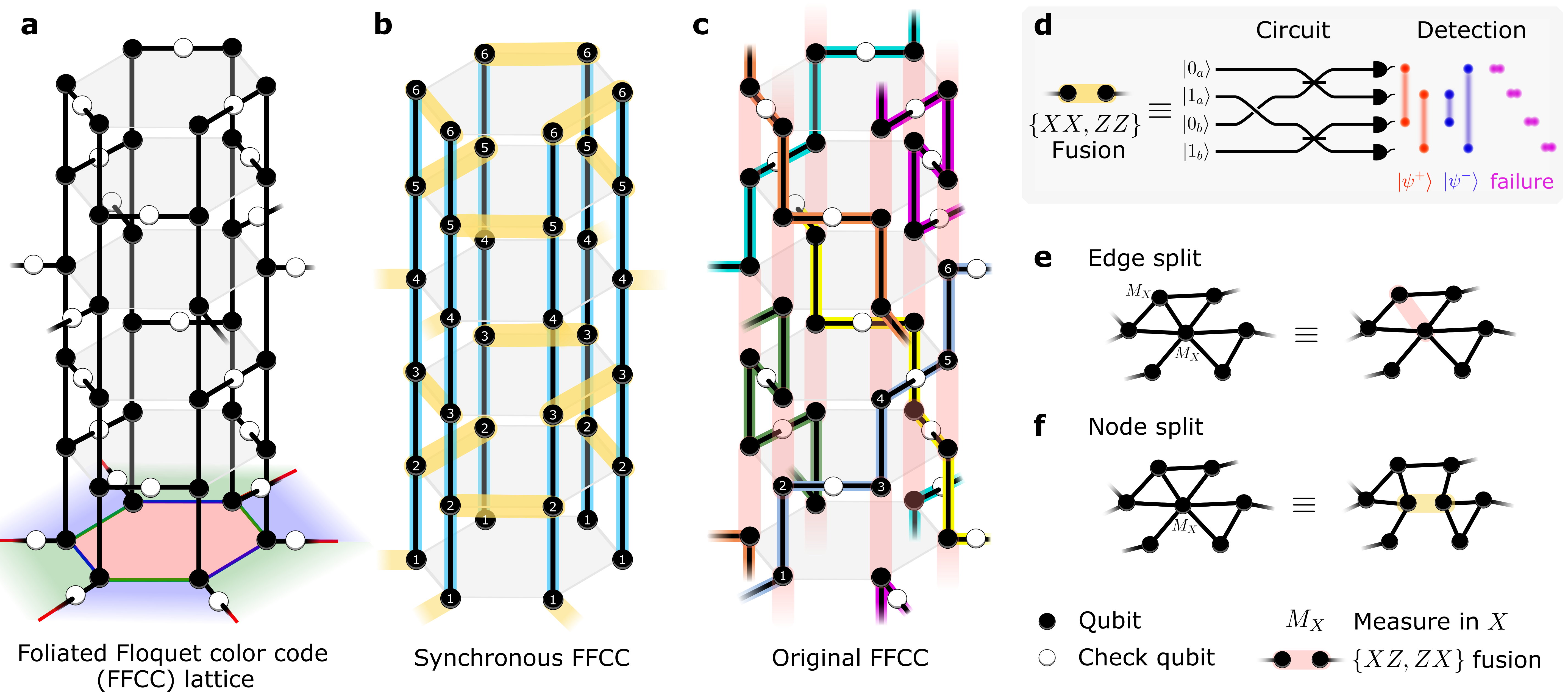}
    \caption{\textbf{Decomposition of FFCC lattice.}  \textbf{(a)} Cluster state for FFCC lattice. In the measurement-based picture, all qubits are measured in the Pauli-$X$ basis to provide syndrome data for error correction. \textbf{(b)} The lattice is decomposed into a fusion network consisting of six linear cluster states localized in one unit cell, with fusions applied along the horizontal edges. \textbf{(c)} Original FFCC fusion network~\cite{paesani_high-threshold_2022} where resource states (identified by different colors) extend into other unit cells and fusions connect different layers.
    \textbf{(d)} Linear-optic circuit for a type-II $\{XX,ZZ\}$ fusion and possible outcomes based on different detection patterns. Successful fusions project two input qubits into a Bell state $\ket{\psi^\pm}$, while fusion failure leads to single-qubit measurements. 
    \textbf{(e)} Edge split rule is used to decompose the FFCC lattice into \textbf{(c)}. \textbf{(f)} Node split rule is applied to convert a qubit into fusions between two additional qubits. When combined with $X$ measurements on the neighbors, this results in \textbf{(b)}.}.
    \label{fig:combined_lattice}
\end{figure*}

Here, we present a fusion-based architecture tailored to quantum emitters and analyze how relevant physical processes affect its fault-tolerant performance. The architecture is based on the foliated Floquet color code (FFCC)~\cite{paesani_high-threshold_2022} where each resource state can be deterministically generated by a single quantum emitter, with high tolerance to photon loss and qubit errors. 
The fault-tolerant performance of FFCC was previously analyzed using a phenomenological error model for fusions, in which erasure and error occurs independently for each fusion measurement~\cite{paesani_high-threshold_2022}, without considering physical noises dominant in quantum emitters.
We develop a framework to describe how these noises propagate to the logical qubit level of the architecture and evaluate its performance in terms of fault-tolerant thresholds.
Specifically, we investigate photon distinguishability, Markovian spin noise in spin-photon interfaces, and photon loss.
We show that our tailored architectures are highly resilient to these noises, as summarized in Table~\ref{tab:threshold}.

\begin{table}
    \centering
    \begin{tabular}{|c|c|c||c|c|c|c|}
        \hline
        \textbf{Lattice} & \makecell{\textbf{Fusion}\\ \textbf{type}}{} & \makecell{\textbf{Code}\\ \textbf{size}}{} & $\ell$ & $1-V$ & $p^{(s)}$ & $p^{(s)}_Z$ \\
        \hline
        sFFCC & REP & $m=5$ & $2.29\%$ & $4.5\%$ & $0.125\%$& $0.19\%$ \\
        sFFCC & RUS & $N=10$ & $8\%$ & $4\%$ & $0.15\%$ & $0.59\%$ \\
        \hline
    \end{tabular}
    \caption{\textbf{Summary of main results}. sFFCC: synchronous foliated Floquet color code lattice. REP: encoded fusion with repetition code of fixed code size $m$. RUS: repeat-until-success encoded fusion with varying code size $m\leq N$ where $N$ is the maximum number of physical fusion attempts. $\ell$, $V$ and $p^{(s)}$/$p^{(s)}_Z$ are thresholds for photon loss, photon indistinguishability from different emitters, and spin depolarizing/$Z$ error probability, respectively.}
    \label{tab:threshold}
\end{table}

The paper is structured as follows. In Sec.~\ref{sec:lattice} we introduce the FFCC lattice and rules to decompose the lattice into fusions between linear photonic cluster states.
Next, in Sec.~\ref{sec:rsg} we revisit the protocol for deterministic generation of the necessary photonic resource states from single-photon emitters. In Sec.~\ref{sec:performance} we then show the baseline fault-tolerant properties of the architecture.
Sec.~\ref{sec:physicalerrors} discusses the physical noise mechanisms relevant to the emitters. In particular, we show how these errors propagate through the cluster-state generation protocol and impact the performance of fusion gates.

To suppress these errors, in Sec.~\ref{sec:repcodes} we discuss the use of code concatenation, redundantly encoding each qubit of the linear cluster state and performing \textit{encoded fusions} between them.
We consider two types of encoded fusion, based on static repetition codes (REP) and repeat-until-success gates (RUS), and elucidate the relationship between the two.
The thresholds of our construction to various noise models are then presented in Sec.~\ref{sec:thresholds}, before suggesting directions for future work.

\section{FFCC lattice and fusion network}
\label{sec:lattice}

In Ref.~\cite{paesani_high-threshold_2022}, the authors introduced the foliated Floquet color code (FFCC) lattice, constructed by foliating the Floquet color code~\cite{FFC_davydova, Kesselring2024}. The code can be understood both in the circuit based model and by looking at the graph state version of it in Fig.~\ref{fig:combined_lattice}a. Here we describe it in terms of the latter.
In general, a graph state is associated with a certain graph, where each vertex represents a qubit prepared in the state $\ket{+}$ and edges represent $CZ$ gates.
In the Floquet color code, data qubits are situated at the vertices of a hexagonal lattice, and check operations are determined by a coloration of the lattice. 
All faces are assigned a color, such that no two adjacent faces share the same color.
Edge colors are determined by the color of the faces at either end of an edge, e.g., a green edge connects two green faces.
The check operators of the Floquet color code are built up from two-body check measurements defined by the colored edges~\cite{Kesselring2024}.

These two-body measurements are performed in layers corresponding to each color, i.e. first green, then red followed by blue. 
Many rounds of measurement of a given color can be combined into a stabilizer check of the color code, with a +1 eigenvalue in the error free case.
This in-plane connectivity is contained in the  additional \textit{check qubits} (white circles in Fig.~\ref{fig:combined_lattice}a) which are coupled to the data qubits (black circles) involved in the check~\cite{paesani_high-threshold_2022}. A qubit can be encoded into the cluster state and propagates e.g., from the bottom to the top layer by measuring all qubits in the lower layers. From the measured check and data qubits we can then extract syndrome information, which allows us to correct errors. 

\subsection{Fusion-based quantum computation}
\label{subsec:fusion-based}
To construct a fusion-based architecture out of such a topological cluster state, the cluster state is first decomposed into an arrangement of smaller entangled resource states and fusion measurements, called a \textit{fusion network}~\cite{bartolucci_fusion-based_2023}.
A \textit{fusion} is a two-qubit entangling measurement, introduced as a method of building cluster states in Ref.~\cite{Browne2005}.
Type-II fusion gates, which we consider in this work, perform destructive joint Pauli measurements on two qubits. 
For input qubits $a$ and $b$, a fusion ideally measures the commuting operators $X_a X_b$ and $Z_a Z_b$~\cite{bartolucci_fusion-based_2023} with Pauli operators $X/Z$, thereby projecting them into a Bell state. This is referred to as a $\{XX,ZZ\}$ fusion.

Here we are interested in optical implementations of fusion networks. Photonic qubits can be represented by two orthogonal modes encoded, e.g., in polarization, frequency, time-bin and photon number. 
Dual-rail encoding, which uses the presence of a photon in either of two orthogonal modes as a qubit, offers loss-tolerance properties which make it appealing for FBQC~\cite{bartolucci_fusion-based_2023}.
With linear optics, fusion measurements can be implemented on dual-rail qubits by a simple four-mode circuit, with photon counting detectors on each mode (Fig.~\ref{fig:combined_lattice}d). 
This linear circuit performs fusion measurement with 50\% success probability. When the fusion fails, measuring single qubit operators $Z_{a}$ and $Z_{b}$ retrieves the $Z_a Z_b$ outcome.
Single qubit gates can be used to rotate the bases of these measurements, e.g., a Hadamard on one of the input qubits results in a $\{XZ,ZX\}$ fusion, and Hadamards on both inputs lead to a $\{ZZ, XX\}$ fusion which retrieves $XX$ upon fusion failure.

Fusion networks can be conveniently described in the \textit{stabilizer formalism}~\cite{Gottesman1997}, where quantum states are represented as eigenstates with eigenvalue $+1$ of a set of operators, e.g., the graph state in Fig.~\ref{fig:combined_lattice}a is the eigenstate of a set of operators $X_i\prod_{j\in \mathcal{N}_i} Z_j$, where $i$ is the qubit index, and $\mathcal{N}_i$ is the neigborhood of qubit $i$ which refer to all qubits connected to it by an edge in the graph. In the fusion-based picture, fusion measurements and resource states can both be represented by subgroups of the Pauli group, denoted $\mathcal{F}$ and $\mathcal{R}$, respectively.
$\mathcal{F}$ is the set of operators measured by all fusions, and $\mathcal{R}$ is the stabilizer group of all resource states~\cite{bartolucci_fusion-based_2023}.
The intersection of these two groups gives the \textit{surviving stabilizer group} $C:=\mathcal{R}\cap\mathcal{F}$~\cite{bartolucci_fusion-based_2023}, a group of check operators that return a $+1$ eigenvalue in the absence of errors.
Errors in the resource states or fusion measurements that anti-commute with elements of $C$ result in negative eigenvalues for one or more check operators.
The set of eigenvalues of check operators is called the syndrome, and is interpreted by a decoder for error correction in FTQC.
\subsection{Constructing fusion networks from lattice decomposition}
Our method for constructing a fault-tolerant fusion network is to decompose a cluster state into simpler resource states using the concepts of \textit{edge split} and \textit{node split}.

An \textit{edge split} rule is constructed by noting that a $\left\{XZ, ZX\right\}$ fusion is equivalent to a $CZ$ gate between the qubits involved in the fusion, before both qubits are measured in the $X$ basis (Fig.~\ref{fig:combined_lattice}e).
For any edge in a cluster state, we can apply an edge split, introducing a fusion measurement, and assert that a $\{XZ, ZX\}$ fusion will recover the same stabilizer group as the cluster state protocol, up to Pauli frame updates due to the measurement outcomes.
%
%

For \textit{node splits}, an equivalence is identified between the measurement of a cluster state node in the $X$ basis and a $\left\{XX, ZZ\right\}$ fusion between two nodes that are introduced at the same site.
One of the new nodes is connected to a subset of the neighbours of the original node, with the other connected to the complement, as shown in Fig.~\ref{fig:combined_lattice}f. A similar concept of node splitting was introduced in Ref.~\cite{nickerson_measurement_2018} but not in the context of fusions.

The proofs for both rules are provided in Supplementary Note~\ref{supp:proof}. In general, these decomposition rules can be systematically applied to any cluster state until target resource states are obtained.
Any remaining qubits that do not take part in fusion gates have implicit $X$ measurements on them, and so we follow Refs.~\cite{bartolucci_fusion-based_2023, paesani_high-threshold_2022} and directly prepare the resource states where these qubits have been projected onto $\ket{+}$.

To obtain the synchronous FFCC fusion network in Fig.~\ref{fig:combined_lattice}b from decomposition of the FFCC lattice, we first apply a node split on each check qubit of the lattice (white circles in Fig.~\ref{fig:combined_lattice}a) and replaces it with two qubits, which are measured by a $\{XX,ZZ\}$ fusion. 
Next, since the two qubits (black) that were connected to the check qubit prior to the node split do not participate in fusions, we consider them as ``virtual qubits" that are not generated by actual hardware, but are rather assumed to be measured in the $X$ basis resulting in $+1$ outcomes. $X$ measurements on these qubits remove them from the graph and have their neighbors connected, then add Hadamard rotations on the neighboring qubits which participate in the $\{XX,ZZ\}$ fusion (Supplementary Note~\ref{supp:proof}). As such, the fusion failure basis is switched to $XX$. 
%
This results in a fusion network consisting of linear cluster states shown in Fig.~\ref{fig:combined_lattice}b, which we refer to as ``synchronous FFCC'', where qubits on the same layer can be simultaneously generated.
As we shall describe later, this synchronicity allows for the use of RUS fusions, providing significant practical advantages for the implementation and noise tolerance of this scheme. 
As a comparison, the original FFCC fusion architecture (Fig.~\ref{fig:combined_lattice}c) was constructed using edge splits.

\begin{figure}
    \centering
    \includegraphics[width=\columnwidth]{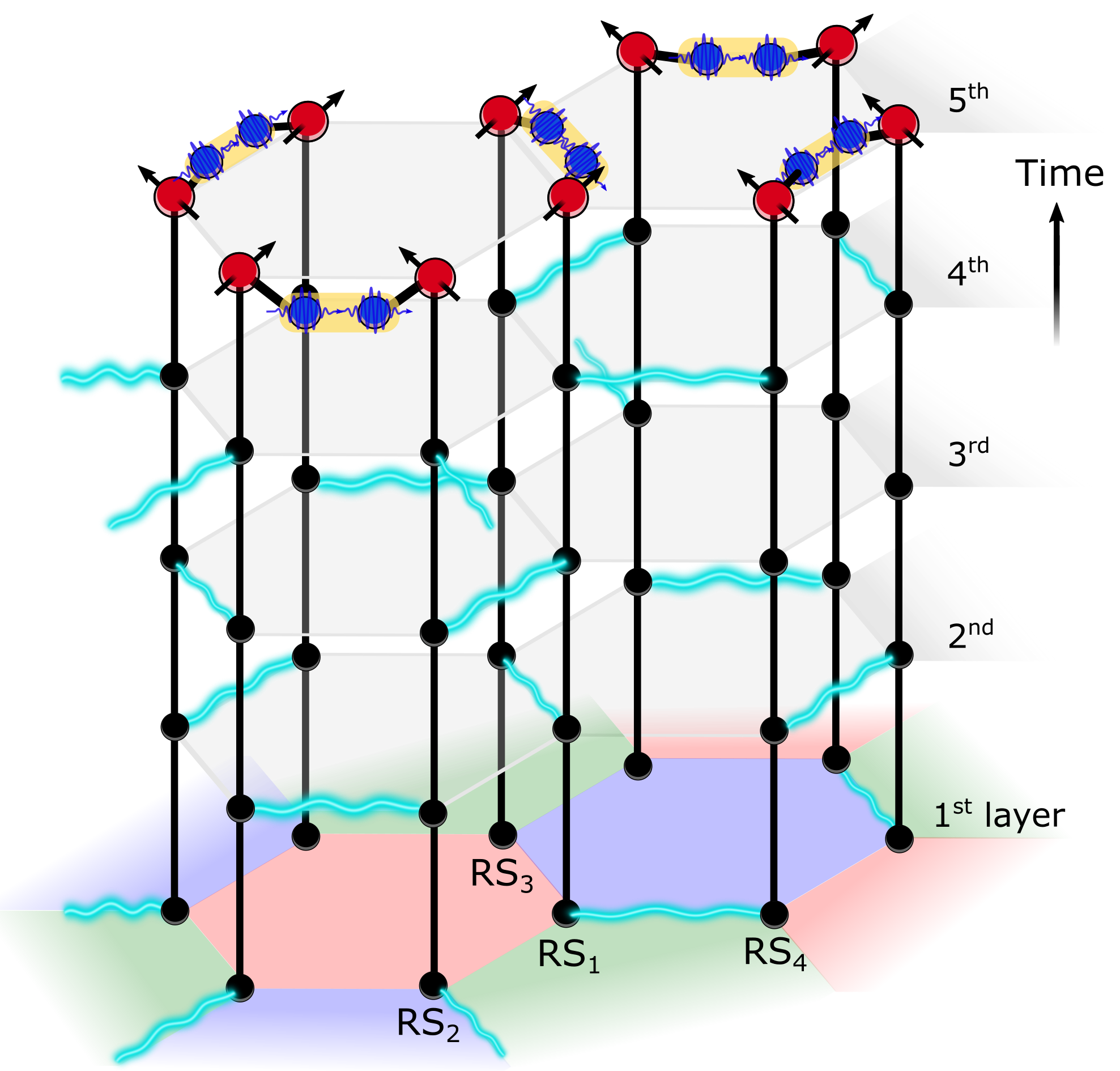}
    \caption{\textbf{Fusion network for synchronous FFCC}. Resource states (RS) are generated by an array of quantum emitters (red circles). Fusions (yellow) are performed between photons to obtain pair-wise parity checks (cyan wavy lines) layers by layers.
    }
    \label{fig:network}
\end{figure}

Another important advantage of the synchronous FFCC network is that resource states remain spatially local and interact with only three neighbouring states. For resource states generated by quantum emitters, this feature ensures that routing requirements for photons from different emitters remain constant with the size of the computation (Fig.~\ref{fig:network}), in contrast to the original scheme.

\section{Resource state generator}
\label{sec:rsg}

\begin{figure}[h]
    \centering
    \includegraphics[width=\columnwidth]{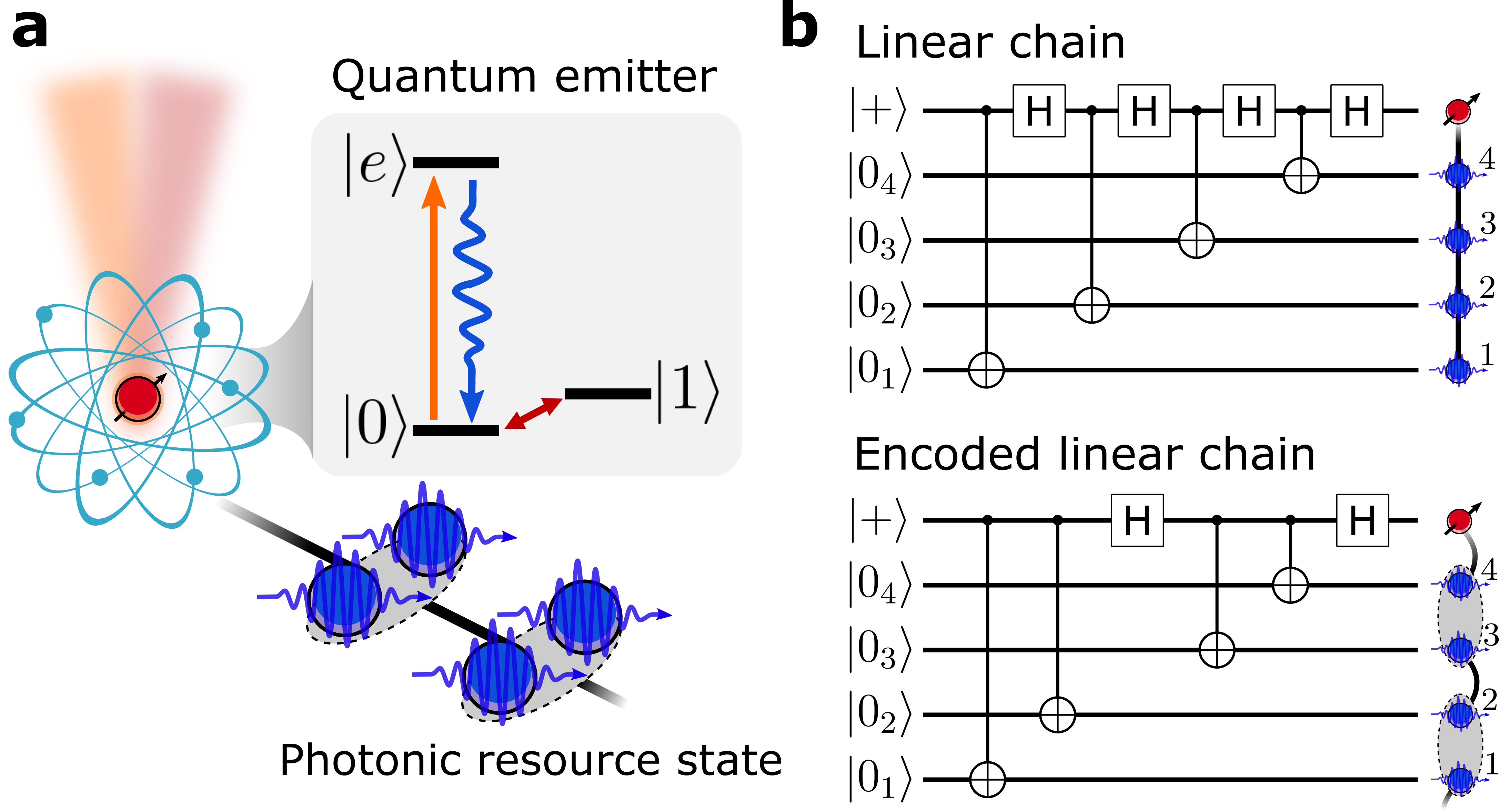}
    \caption{\textbf{Generation of photonic resource states with a quantum emitter}. 
    \textbf{(a)} Schematic of one possible energy level structure of the emitter to generate time-bin photon entangled with its spin. A photon is emitted (blue wavy arrow) upon optical excitation (orange) of the $\ket{0}\to\ket{e}$ cycling transition. A time-bin encoded photon is generated by two optical excitations of the transition interleaved with a spin $\pi$-rotation~\cite{Lee2019}. Hadamard gates are implemented by coherently controlling the emitter ground states (red). \textbf{(b)} Quantum circuits to generate linear and encoded linear chain states.} 
    \label{fig:RSG}
\end{figure}

In Ref.~\cite{lindner_proposal_2009}, Lindner and Rudolph proposed a simple protocol for the generation of photonic resource states by utilizing spin states of a single quantum emitter and photon emissions (Fig.~\ref{fig:RSG}a). 
The protocol requires a quantum emitter capable of generating single photons on-demand, where photon emission is conditioned on the state of the emitter.
This is possible using, e.g., an $\Lambda$-type level structure (Fig.~\ref{fig:RSG}a) or a four-level system with degenerate closed optical transitions~\cite{lindner_proposal_2009}, with emitter spin ground states $\ket{0}$ and $\ket{1}$. Hadamard gates are applied to the spin to coherently transfer from $\ket{0}$ ($\ket{1}$) to $\ket{+}$ ($\ket{-}$), where $\ket{\pm}=(\ket{0}\pm\ket{1})/\sqrt{2}$.

To generate an $n$-photon linear cluster state, the protocol proceeds as:
\begin{enumerate}
    \item Prepare the emitter in the superposition state \\ $\ket{\psi} = \ket{+} = 1/\sqrt{2} \left(\ket{0} + \ket{1}\right)$.
    \item Repeat the following  steps $n$ times:
    \begin{enumerate}
    \item Optically excite the emitter to emit a photon in state $\hat{a}_{0}^{\dagger}\ket{\emptyset}$ or $\hat{a}_{1}^{\dagger}\ket{\emptyset}$ entangled with the spin states: \\ $\ket{\psi} \rightarrow 1/\sqrt{2} \left(\ket{0}\hat{a}_{0}^{\dagger} + \ket{1}\hat{a}_{1}^{\dagger} \right)\ket{\emptyset}$;
    \item Perform a Hadamard gate on the spin: \\ $\ket{\psi} \rightarrow 1/\sqrt{2} \left(\ket{+}\hat{a}_{0}^{\dagger} + \ket{-}\hat{a}_{1}^{\dagger} \right)\ket{\emptyset}$;
    \end{enumerate}
    \item Measure the spin in the computational basis.
\end{enumerate}
This prepares the $n$-photon linear cluster state up to a $Z$-rotation on one of the photons, dependent on the measurement outcome.
Depending on the choice of qubit encoding, there are different approaches to implement Step 2a. For time-bin encoding (Fig.~\ref{fig:RSG}a), it is equivalent to applying two optical excitation pulses separated by a spin $\pi$-rotation pulse~\cite{Lee2019,tiurev_fidelity_2021}. 
An equivalent quantum circuit for a linear cluster state or chain is shown in Fig.~\ref{fig:RSG}b.
In this picture, the protocol proceeds via a sequence of Hadamard and controlled-NOT gates, where the emitter acts as the control qubit and the target (photonic) qubits are initialized in $\ket{0}$.

An encoded linear chain state can also be generated by a single emitter, where each qubit is redundantly encoded in $m$ photons~\cite{Hilaire2023neardeterministic}. As shown in Fig.~\ref{fig:RSG}b, this is done by omitting some interstitial Hadamard gates on the emitter in Step 2b of the circuit, which generates a linear chain with each encoded qubit (grey dashed circle) in an $m$-photon repetition code~\cite{Hilaire2023neardeterministic}. In the graph state picture these additional photons can be described as branches on the linear chain. This encoding is essential for improving the loss and fault tolerances of the fusion-based architecture, as we shall cover in Sec.~\ref{sec:repcodes}.

\section{Fault-tolerance under measurement errors}
\label{sec:performance}

Before incorporating physical noises from quantum emitters into the architecture, we first investigate the fault-tolerant properties of the synchronous FFCC fusion network, where a phenomenological noise model is introduced for fusion measurements. In this model independent and identically distributed (i.i.d.) errors and erasures occur on each fusion outcome with probabilities $p_{\text{err}}$ and $p_{\text{eras}}$ respectively.

For synchronous FFCC, each fusion outcome contributes to two checks, for the primal and dual lattice, respectively~\cite{paesani_high-threshold_2022}. From here a syndrome graph can be constructed. It has a degree of $12$, with each check consisting of six $XX$ and six $ZZ$ outcomes. 
The decoding process consists of two steps: First, for lost measurement outcomes due to fusion erasures, we adapt methods outlined in Ref.~\cite{barrett_fault_2010} to merge checks and construct \textit{supercells}, which have higher weight but depend only on measurements that have not been erased. 
The syndrome graph is then modified with the newly merged cells, and a new logical correlation operator is found~\cite{paesani_high-threshold_2022}. 
If such an operator cannot be found, this leads to a \textit{logical erasure}.
Second, with the modified syndrome graph, we use the minimum-weight perfect matching (MWPM) decoder implemented with \textit{Pymatching}~\cite{higgott2023sparse} to find a minimum weight correction to the given syndrome.
If the correction results in a flipped logical correlation operator, this constitutes a \textit{logical error}.

\begin{figure}
    \centering
    \includegraphics[width=0.9\columnwidth]{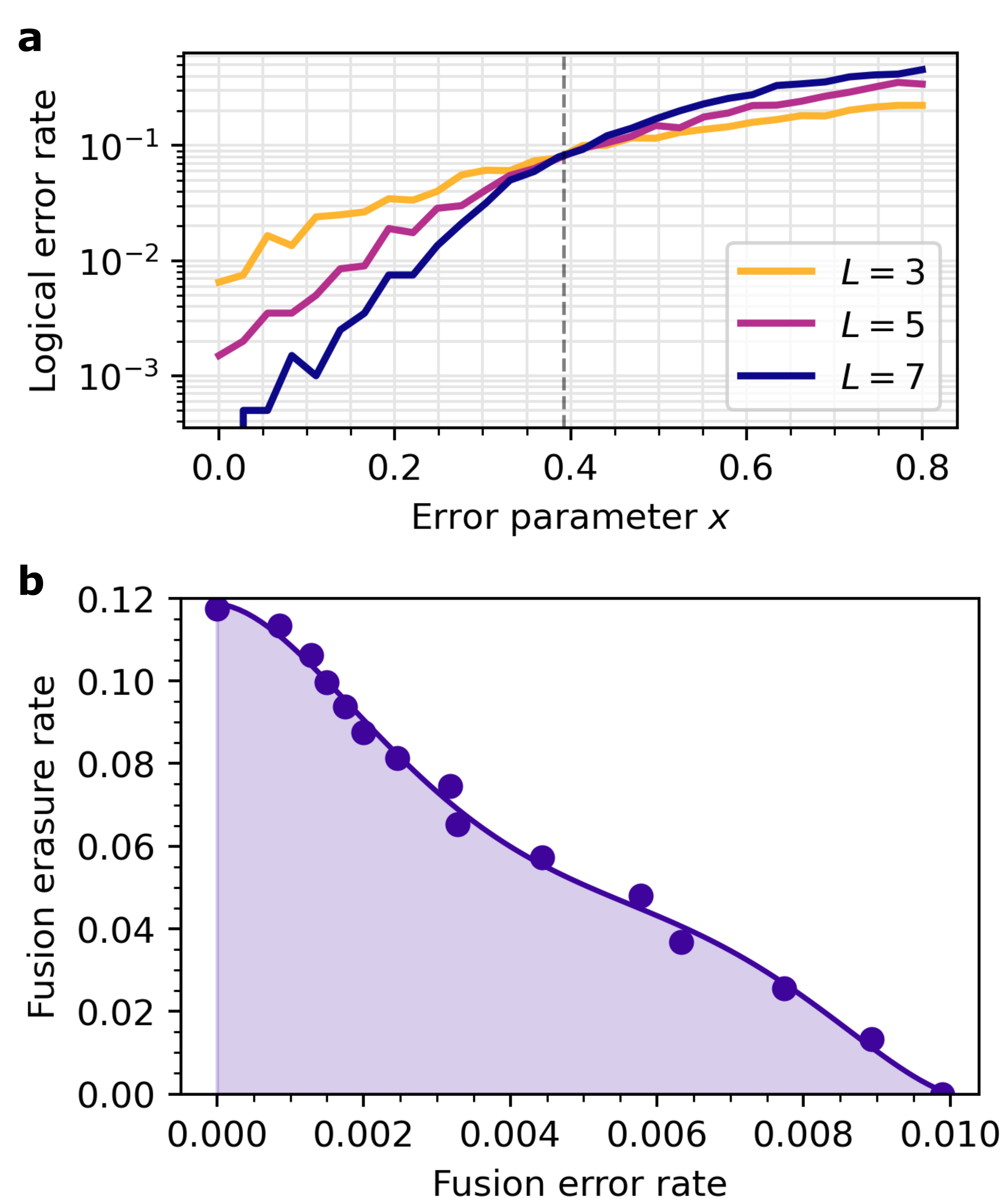}
    \caption{\textbf{FFCC lattice performance.} \textbf{(a)} An example of threshold estimate by sweeping $x$ to change the fusion error and erasure rates, with $p_{\text{err}}=0.0027+0.014x$ and $p_{\text{eras}}=0.0148+0.0082x$. The logical error rate is computed for various lattice sizes $L$. The threshold $x^*=0.392$ corresponds to $(p_{\text{err}},p_{\text{eras}})=(0.8\%,1.8\%)$. \textbf{(b)} Shaded region represents the fault-tolerant region for synchronous FFCC fusion-based architecture, where errors can be reliably corrected. The solid curve is a polynomial fit to guide the eye. }
    \label{fig:lattice_threshold}
\end{figure}

Simulations are performed by Monte-Carlo sampling of different configurations of fusion error and erasure.
For each data point, $N_{\text{trials}}=10^4$ trials are used to estimate the logical error rate $R=N_{\text{fail}}/N_{\text{trials}}$, where $N_{\text{fail}}$ is the number of trials when the decoding fails due to either logical errors or erasures. The lattice is simulated using periodic boundary conditions, as we are interested primarily in the behaviour of the bulk. To identify the region of fault rates that can be corrected by the architecture, we linearly parameterise $p_{\text{err}}$ and $p_{\text{eras}}$ by an error parameter $x$, and calculate logical error rates for various lattice sizes $L$ as $x$ is varied.

Fig.~\ref{fig:lattice_threshold}a shows an example of simulated logical error rate as a function of lattice size and error parameter $x$, where the threshold occurs at $(p_{\text{err}},p_{\text{eras}})=(0.8\%,1.8\%)$. From the intersection between error curves, we extract the threshold value of the error parameter $x^{*}$, below which errors can be arbitrarily suppressed by increasing the lattice size.
The fault-tolerant region for synchronous FFCC is shown in Fig.~\ref{fig:lattice_threshold}b.
The marginal fusion error threshold is found to be $1\%$, and the marginal fusion erasure threshold is $ 11.9\%$, in line with previous proposals based on similar decompositions~\cite{paesani_high-threshold_2022}. 

\section{Physical error models}
\label{sec:physicalerrors}
In this section we introduce various physical error and loss models into the generation of resource states with quantum emitters, which give more insights on the fault-tolerant properties of the synchronous FFCC fusion architecture in a realistic implementation.

\subsection{Photon loss and distinguishability between emitters}
Photon loss is the dominant failure mode of photonic devices, and can be caused by imperfect photon extraction from the quantum emitter, propagation loss in optical elements, and photo-detector inefficiency.
For a type-II fusion gate, loss of either of the two input photons (Fig.~\ref{fig:physical_errors}a(i)) results in the erasure of both measurement outcomes.
Importantly, because the type-II fusion requires two detections, this erasure is heralded.
Here we bundle all forms of loss into one parameter $\eta$ which represents the end-to-end efficiency.
To the first order, a finite multi-photon component  $g^{(2)}>0$  in one of the input photons also leads to erasure of both measurement outcomes, as it would cause $(n>2)$-fold coincidence detection events in the fusion.
In this work we neglect the effect of this error as we expect it to be less prevalent than photon loss, but as tolerable loss rates are increased, it will be important to incorporate second-order effects induced by the combination of $g^{(2)}>0$ and loss.

For FBQC, high-fidelity quantum interference between photons from different emitters is required, since photon distinguishability limits the Hong-Ou-Mandel (HOM) interference~\cite{Hong1987} required to perform successful type-II fusion gates.
For imperfect quantum emitters in an inhomogeneous magnetic environment, we may expect the spectral profile of emitted single photons to be time varying, or mismatched between two sources, such that interfering photons are \textit{partially distinguishable} (Fig.~\ref{fig:physical_errors}a(ii)).
For a $\{XX, ZZ\}$ fusion, partial distinguishability causes a $Z$ error on the successfully projected Bell state with rate $\frac{1-V}{4}$, where $V$ is the HOM dip visibility~\cite{sparrow2018_phdthesis, Rhode}. 
Fusion error and erasure rates due to photon distinguishability are calculated analytically, and verified by simulations using the one-bad-bit model~\cite{sparrow2018_phdthesis}.
It is important to note that for simplicity we have not considered photon distinguishability from the same emitter. In principle, a finite distinguishability induces measurement errors in $ZZ$ outcomes obtained from fusion failure.

\subsection{Spin errors during resource state generation}\label{sec:rsg_spin_error}

\begin{figure}
    \centering
    \includegraphics[width=\columnwidth]{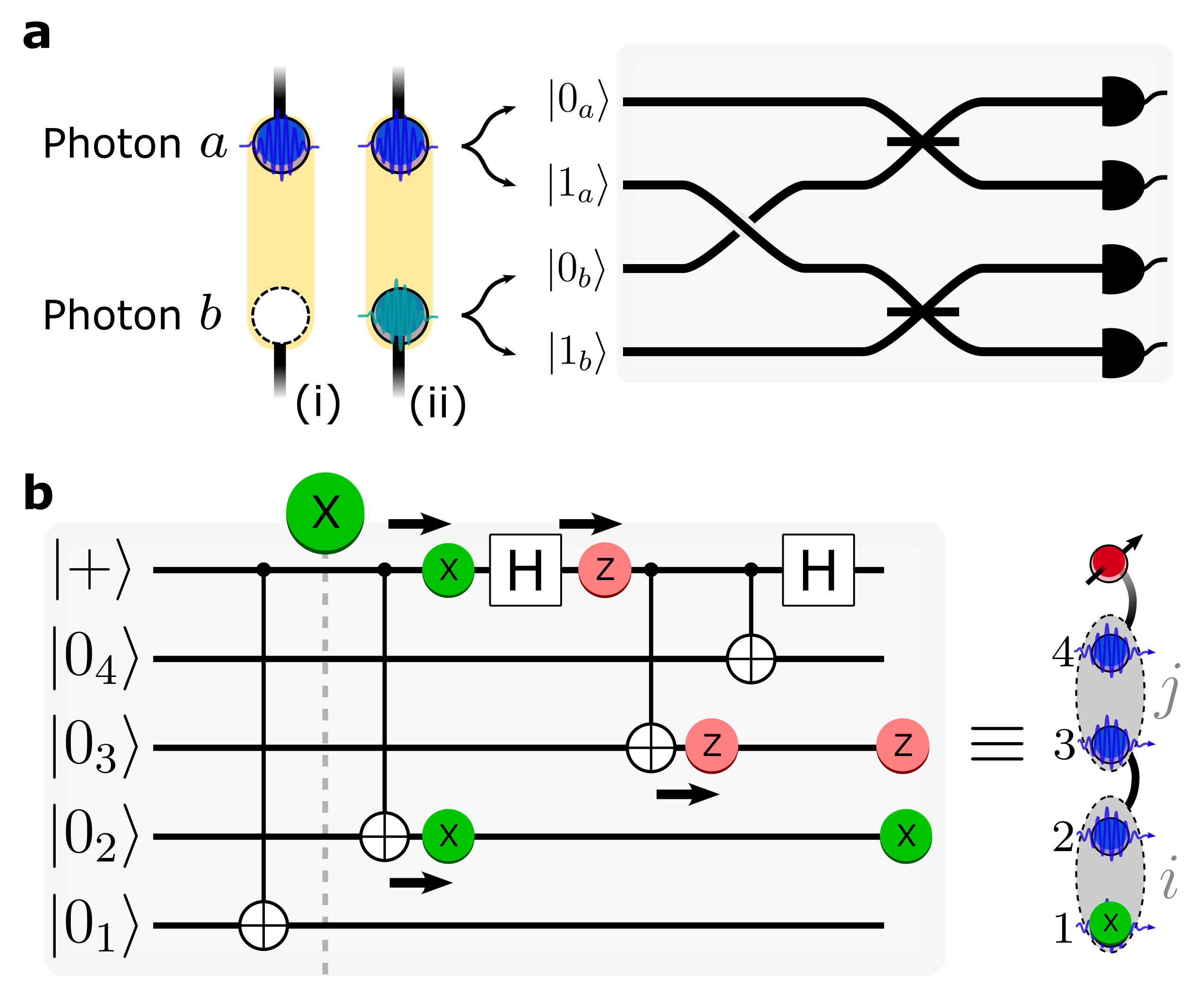}
    \caption{\textbf{Physical noises considered in the architecture analysis.} \textbf{(a)} Linear-optic type-II fusion with photon noises. (i) Loss in one of the photons leads to fusion erasure. (ii) Distinguishable photons generated by the emitter do not interfere and lead to both fusion erasure and errors. \textbf{(b)} Example of spin error propagation into photon errors in the generation of redundantly encoded linear cluster states. A spin-flip error (denoted by $X$ gate, dashed line) in the circuit leads to a $X$ error on photon 2, and a $Z$ error on photon 3, which is equivalent to having a $X$ error on photon 1, up to stabilizers.}
    \label{fig:physical_errors}
\end{figure}

During resource state generation, spin-flip (denoted by $T_1$  time) or dephasing (spin coherence time $T_2$) errors may arise owing to an imperfect quantum emitter, which then propagate through the generation circuit and manifest as photonic errors in the resulting state.
The noise accumulated during resource state generation may have correlations that are not well captured by an i.i.d. error model, so circuit-level noise modelling is important.

We model spin noise with a depolarizing channel where stochastic single-qubit Pauli $X$, $Y$ and $Z$ errors are applied to the emitter spin during state generation, with probabilities $p^{(s)}_X=p^{(s)}_Y=p^{(s)}_Z$.
In general, the relaxation and dephasing of a quantum emitter due to its coupling to a Markovian noise bath can be modelled by a Pauli error channel~\cite{lindner_proposal_2009,deGliniasty2024spinopticalquantum}, where $p^{(s)}_X=p^{(s)}_Y\neq p^{(s)}_Z$ with $T_2 \leq 2 T_1$. 
In the simulation, we sample spin errors from the Pauli group $\mathcal{P} = \{I, X, Z, Y\}$ at each step during state generation with probabilities $\{1-p^{(s)}, \frac{p^{(s)}}{3}, \frac{p^{(s)}}{3}, \frac{p^{(s)}}{3}\}$, where $p^{(s)}$ is the depolarizing error probability. 
Therefore, for a Markovian noise model we have $\frac{p^{(s)}}{3}=\frac{1}{4}\big[1-\exp(-\tau/T_2)\big]$~\cite{deGliniasty2024spinopticalquantum}
where $\tau$ is the time required to generate a photon. We note that a depolarizing channel, where $T_2 = T_1$, is a worst-case assumption for the noise allowing us to infer a lower bound on the threshold for $T_2$ from $p^{(s)}$.

A mapping from spin to photonic errors is constructed by propagating these errors through the circuit using the identities $C_{\text{NOT}}\left(X \otimes I \right) C_{\text{NOT}}^{-1} \equiv X\otimes X$,  $C_{\text{NOT}}\left(Z \otimes I\right) C_{\text{NOT}}^{-1} \equiv Z\otimes I$ and $HXH \equiv Z$, where the spin (each photon) is initialized in $\ket{+}$ ($\ket{0}$).
The $C_{\text{NOT}}$ gate ``copies" $X$ errors from the spin onto the photon, whereas $Z$ spin errors remain as single qubit errors.
A Hadamard gate exchanges the error type and prevents $X$ errors from proliferating across the whole resource state.
By considering the stabilizer group of the resource state, a set of photon errors on the resulting state can be obtained, as two sets of Pauli operators that are equivalent up to stabilizers will give the same stabilizer measurement outcomes~\cite{Gottesman1997}.
Given that the photon is initialized in $\ket{0}$, a $Z$ spin error before a $C_{\text{NOT}}$ is equivalent to a single $Z$ errors on the emitted photon.
Similarly, a $Y$ spin error before the $C_{\text{NOT}}$ is converted into a $X$ spin error and a $Y$ error on the generated photon~\cite{lindner_proposal_2009}. 

As an example, we now consider error propagation in a redundantly encoded linear cluster state (Fig.~\ref{fig:physical_errors}b), which is the resource state for synchronous FFCC. 
Here a spin $X/Y$ error generally leads to a mixture of photon $X/Y$ and $Z$ errors, depending on when the error occurs.
For instance, a spin $X$ error occurred after the first emitted photon will lead to $X$ errors on all subsequently emitted photons of the current branch $i$, as well as a $Z$ error on the first photon emitted in the next branch $j$.
Therefore, a $X$ spin error appeared in one branch is correlated with a $Z$ photon error in the next generated branch.
In Fig.~\ref{fig:physical_errors}b, since the encoded linear cluster state $\ket{\psi_{\text{enc}}}$ is stabilized by the generator $\bar{X}_i\bar{Z}_j= X_1 X_2 Z_3$ with a local repetition code described by Eq.~(\ref{eq:rep_code_stab}), we have $X_2 Z_3\ket{\psi_{\text{enc}}} = X_2 Z_3\bar{X}_i\bar{Z}_j \ket{\psi_{\text{enc}}}= X_1 \ket{\psi_{\text{enc}}}$, equivalent to applying a $X$ error on photon $1$.

\section{Local repetition codes}
\label{sec:repcodes}

Code concatenation is frequently used in quantum error correction to suppress errors, at the price of increased qubit count and more complex decoding schemes.
In Sec.~\ref{sec:performance} we have shown that our baseline architecture can tolerate moderate levels of hardware imperfection. Here we explore whether we can increase the robustness of this architecture by employing a simple code concatenation scheme for the resource state. 

Specifically we consider a redundantly encoded linear cluster state, where each qubit is encoded in an $m$-photon $Z$-repetition code, with logical operators and stabilizers $\mathcal{S}$ of the local code given by
\begin{equation}
    \label{eq:rep_code_stab}
    \mathcal{S} = \langle Z_{1}Z_{j} \rangle_{j=2}^{m}, \qquad \bar{X} = X^{\otimes m}, \qquad \bar{Z} = Z_{1}.
\end{equation}
This code is natively generated on each qubit of the linear cluster state by the resource state generation protocol described in Sec.~\ref{sec:rsg} by omitting Hadamard gates.

\begin{figure}
    \centering
    \includegraphics[width=\columnwidth]{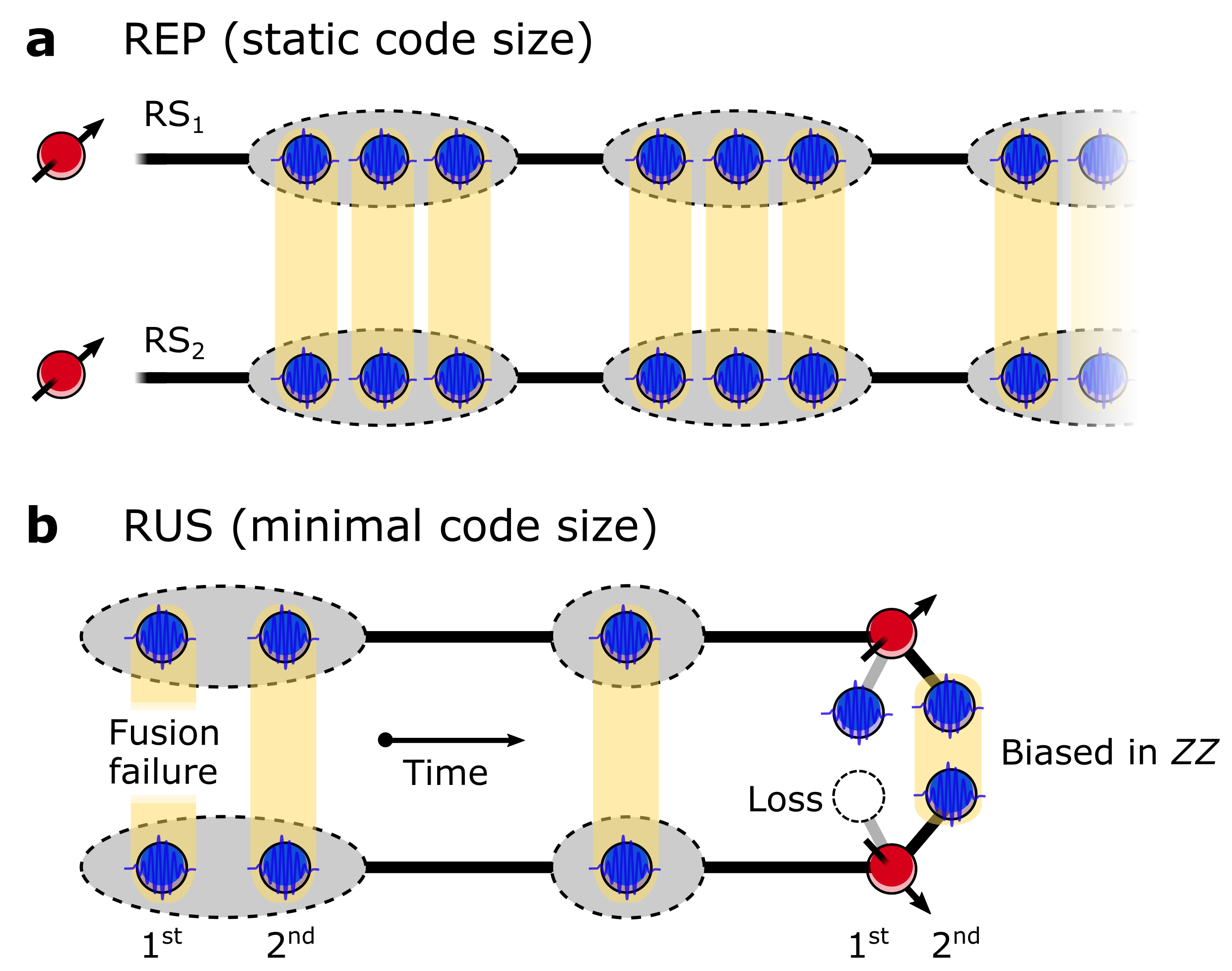}
    \caption{\textbf{Encoded fusion protocols.} \textbf{(a)} With a static repetition (REP) code size of $m=3$, an encoded fusion equates to applying $m$ transversal physical fusions between encoded qubits of each resource state. \textbf{(b)} Repeat-until-success (RUS) encoded fusion performs a minimal number of physical fusions to recover one or two measurement outcomes. If the first physical fusion fails, another pair of photons is emitted for a second attempt. If one of the photons is lost, the next fusion will be biased in $ZZ$~\cite{sahay_tailoring_2022} to guarantee an outcome in $\overline{ZZ}$.}
    \label{fig:logfusion}
\end{figure}

\subsection{Encoded Fusion with repetition codes}
With a local encoding, the physical fusions are substituted for encoded fusion measurements.
For an $m$-qubit repetition code, the fusion group $\mathcal{F}$ associated with an encoded fusion between two qubits $a$ and $b$, is
\begin{equation}
    \label{eq:log_fusion_group}
    \mathcal{F} = \langle \bar{X}_{a}\bar{X}_{b}, \bar{Z}_{a}\bar{Z}_{b} \rangle  \; = \; \langle \bigotimes_{j,k = 1}^{m}X_{a_{j}}X_{b_{k}}, Z_{a_{1}}Z_{b_{1}} \rangle,
\end{equation}
where $a_{j}$ refers to the $j$th physical qubit in the encoded qubit $a$.
These operators can be recovered by performing fusions \textit{transversally} (Fig.~\ref{fig:logfusion}a), such that each qubit from one code undergoes a physical fusion with its equivalent in the other.
For synchronous FFCC, we seek to perform logical $\left\{XX, ZZ\right\}$ fusions on qubits encoded in a $Z$-repetition code of size $m$.
In order to recover the $\overline{XX}$ outcome, all $m$ physical $XX$ outcome must be obtained (Eq.~(\ref{eq:log_fusion_group})), but for $\overline{ZZ}$ a single physical $ZZ$ outcome is sufficient. 
As such, it is beneficial to perform rotated physical fusions on each pair of physical qubits, which measure $XX$ and $ZZ$ on success, and recover the $XX$ outcome upon fusion failure (Sec.~\ref{subsec:fusion-based}).
In this manner, the $\overline{XX}$ outcome is protected by the choice of physical fusion, while $\overline{ZZ}$ is protected by redundancy of the repetition code.

The probabilities of errors and erasures on fusion outcomes using a repetition code with fixed code size (REP) are calculated analytically in Supplementary Note~\ref{supp:RUS}.
We employ a simple majority-voting decoding strategy for errors on the $k$ non-erased $ZZ$ outcomes, such that $\left\lfloor \frac{k-1}{2} \right\rfloor$ errors can be corrected.
The $Z$-repetition code offers no tolerance to $Z$ error, so the error rate on the $\overline{XX}$ outcome, provided it is not erased, is the probability of having an odd number of erroneous $XX$ outcomes.

\subsection{Repeat-until-success encoded fusions}
\label{subsec:RUS_f}
It is well known that adaptivity in physical fusion measurements can greatly improve the recovery probability for encoded fusion outcomes over static methods~\cite{bombin_increasing_2023, Bell_2023_optimizing, Song2024}.
So far we have considered encoded fusions with a fixed number of physical fusions $m$ defined by the repetition code, but a more resource-efficient approach is possible using a repeat-until-success (RUS) fusion.
This can be thought of as a variable-size repetition code, where we only generate the minimal size of code required to achieve an encoded fusion.

The central idea of RUS encoded fusion is to terminate further fusion attempts when one of the following conditions is met: (1) both $\overline{XX}$ and $\overline{ZZ}$ outcomes are recovered from prior attempts, (2) only $\overline{ZZ}$ is recovered, or (3) a maximum number of attempts $N$ is reached.
This reduces the risk of a photon loss erasing the $\overline{XX}$ outcome, and the smaller average code size $m$ reduces the impact of $Z$ errors.
If a loss occurs in a prior fusion attempt, $\overline{XX}$ is no longer recoverable, and the failure basis of the next physical fusion is switched to $ZZ$~\cite{sahay_tailoring_2022} to recover $\overline{ZZ}$ with high probability (Fig.~\ref{fig:logfusion}b).

RUS fusion between two resource states is implemented by an emit-and-measure scheme, where both emitters sequentially emit photons to be fused until one of the above conditions is met.
This means that we only generate the minimum required number of photons and are thus less susceptible to photon loss, but we note that the adaptivity in measurement also comes with higher demands for the hardware, which now needs to condition the control fields on the  measurement outcomes. Furthermore, RUS fusion places scheduling constraints on the order of lattice fusions, which are most simply satisfied by requiring that fusions occur between qubits on the same layer of the lattice (Fig.~\ref{fig:network}). This means RUS fusion is applicable for synchronous FFCC, but is not possible in the original FFCC construction (Fig.~\ref{fig:combined_lattice}c).
\begin{figure}
    \centering
    \includegraphics[width=0.9\columnwidth]{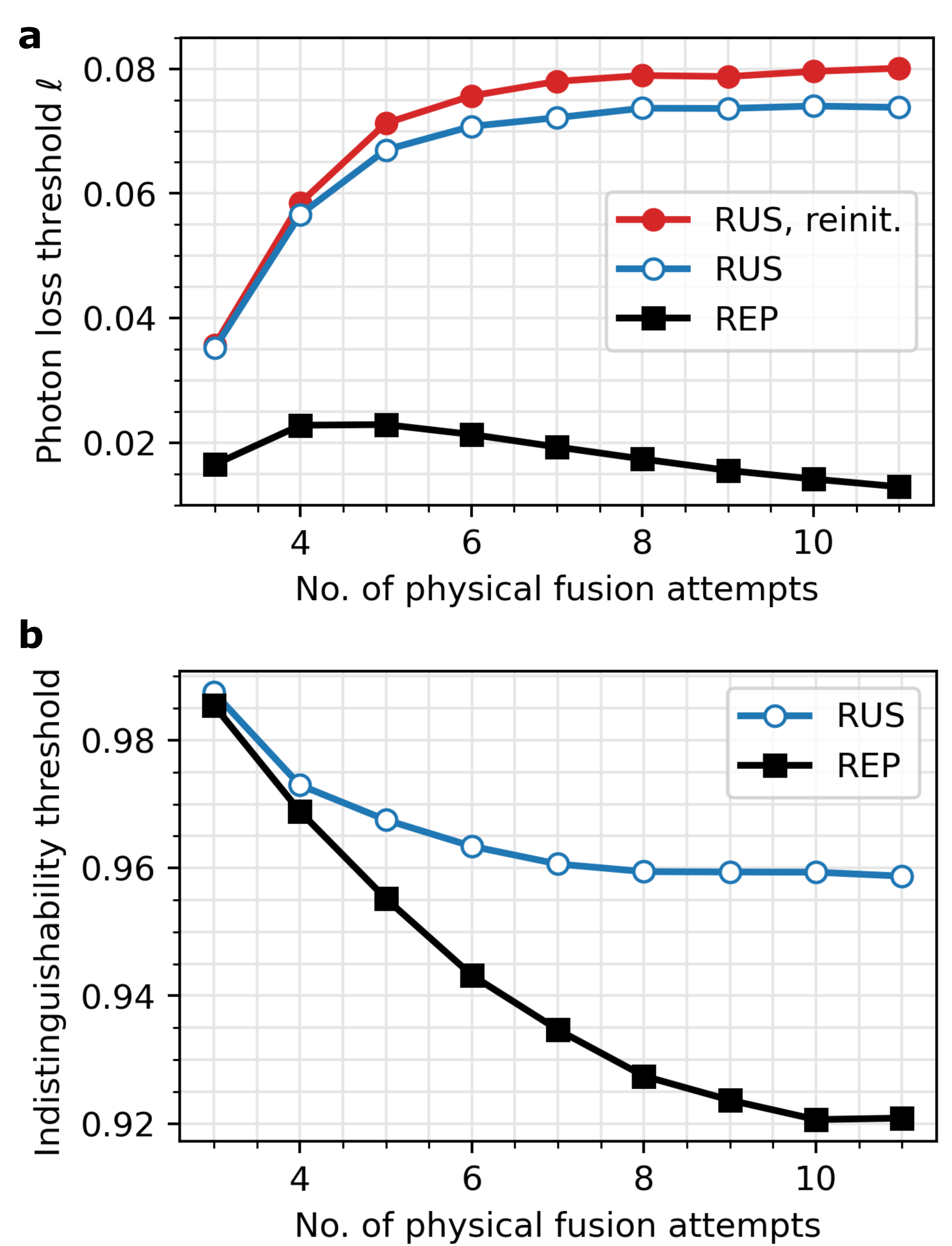}
    \caption{\textbf{Loss and indistinguishability thresholds.} \textbf{(a)} Photon loss threshold as a function of number of fusion attempts per encoded fusion ($m$ for REP and $N$ for RUS). Red solid circle represents RUS with $N$ re-attempts when emitters are re-initialized on the previous layer of fusions. \textbf{(b)} Indistinguishability between photons from different emitters required for fault-tolerance, assuming no photon loss.}
    \label{fig:loss_threshold}
\end{figure}

The probabilities for different encoded fusion outcomes of a single RUS fusion under photon loss and errors can be analytically derived, see Supplementary Note~\ref{supp:RUS}.
Using these analytical expressions, encoded fusion outcomes are sampled to obtain a syndrome graph for calculating fault-tolerant thresholds.

\section{Fault-tolerant Thresholds}
\label{sec:thresholds}
The loss-tolerance thresholds are presented in Fig.~\ref{fig:loss_threshold}a, for various code sizes $m$ for REP and maximal number of physical fusion attempts $N$ for RUS.
The highest REP loss threshold obtained is $\ell=2.29\%$ for $m=5$.
For RUS encoded fusion, we observe that $\ell$ improves significantly as the number of maximally allowed fusion attempts $N$ increases, achieving $\ell=7.4\%$ with $N=10$. $\ell$ can be further improved to $8\%$, by using a strategy where additional $N$ re-attempts of RUS are allowed with both emitters re-initialized from the previous layer of encoded fusions. This trick stems from the observation that a biased physical fusion requires performing single-qubit $Z$ measurements on both photons to generate the $\overline{ZZ}$ outcome. Since the emitted photon and its entangled spin share the stabilizer $ZZ$, measuring the photon in $Z$ effectively projects the spin in $Z$-basis, which re-initializes it. If two emitters were re-initialized from the previous layer of encoded fusions, they are detached from any correlations and hence $N$ additional re-attempts of RUS can be applied to make the next encoded fusion near-deterministic (Supplementary Note~\ref{supp:RUS_inf}).

In the absence of loss, the RUS distinguishability threshold continues to increase with $N$, but saturates at $V=95.9\%$ (Fig.~\ref{fig:loss_threshold}b) when $N=10$, whereas REP can tolerate $V=92\%$ with $m=10$.
The reason for the difference in thresholds is that photon distinguishability between emitters introduces a $X$ error in the rotated fusion gate assumed for these encoded fusions, which is well-tolerated by the repetition code through majority voting.
This strategy is not available using RUS fusions which are terminated once the first $ZZ$ outcome is obtained, thus leading to the difference in thresholds observed here.

For thresholds under spin noise, simulations are performed in two steps: (1) the encoded fusion outcomes are sampled without photon loss. With only fusion failure $P_{\text{fail}}=50\%$, RUS results in either (i) $\overline{XX}$, or (ii) both $\overline{XX}$ and $\overline{ZZ}$ outcomes. (2) For each encoded fusion that ends up in case (i), all physical fusions failed thus $m=N$. For case (ii), the number of fusion attempts $m\leq N$ is sampled according to the probability $(1-P_{\text{fail}})P_{\text{fail}}^{m-1}$. A list of values of $m$ is then obtained and used to extract encoded fusion errors based on a pre-sampled list of photon errors originated from spin noise during resource generation (Sec.~\ref{sec:rsg_spin_error}). It is important to note that the simulation only accounts for spin decoherence of the two emitters involved in the current encoded fusion, whereas other emitters are assumed to be perfectly coherent during the RUS gate.

\begin{figure}
    \centering
    \includegraphics[width=0.9\columnwidth]{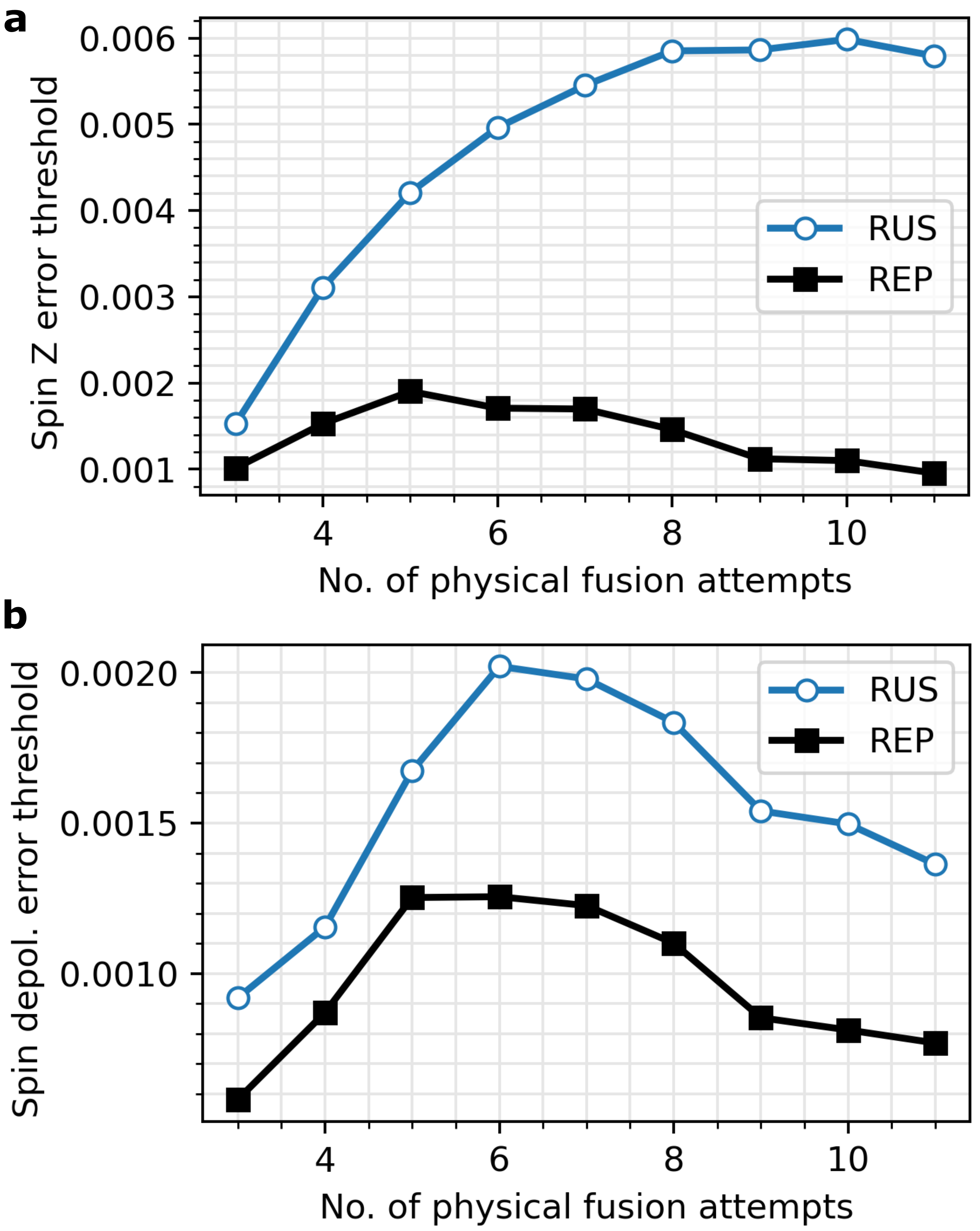}
    \caption{\textbf{Spin error thresholds.} \textbf{(a)} Spin $Z$ error threshold $p^{(s)}_Z$ as a function of number of fusion attempts per encoded fusion ($m$ for REP and $N$ for RUS), with $p^{(s)}_X=p^{(s)}_Y=0$. \textbf{(b)} Spin depolarizing error threshold $p^{(s)}$ where each spin Pauli error probability is identical: $p^{(s)}_X=p^{(s)}_Y=p^{(s)}_Z=p^{(s)}/3$.}
    \label{fig:spin_threshold}
\end{figure}

Fig.~\ref{fig:spin_threshold}a plots the simulated spin $Z$ error threshold with the number of fusion attempts per encoded fusion, with $p^{(s)}_X=p^{(s)}_Y=0$. We observe that for spin errors occurred during the generation of encoded linear chains, the RUS threshold increases up to $p^{(s)}_Z=0.59\%$ at $N=10$ whereas for REP it peaks at $p^{(s)}_Z=0.19\%$ then steadily decreases to $0.095\%$. 
As discussed in Sec.~\ref{sec:rsg_spin_error}, a spin $Z$ error will localize as a $Z$ error in the next emitted photon. For REP the number of fusion attempts $m$ is fixed, such that when $m$ grows the probability of flipping $\overline{XX}$ also increases, and this leads to the decreasing threshold for high $m$. In the limit of low $m$, however, more fusion attempts increases the fusion success probability resulting in a more connected and hence more robust lattice, with an optimal threshold at $m=5$. For RUS, most encoded fusions terminate in a few attempts assuming no photon loss (Supplementary Note~\ref{supp:RUS}), so that only $Z$ errors occurring in the photons emitted during those attempts contribute to a flip in $\overline{XX}$. As a consequence, the spin $Z$ error threshold saturates as we increase $N$.

For spin depolarizing error (Fig.~\ref{fig:spin_threshold}b), both thresholds again increase for low number of fusion attempts thanks to a better connectivity of the graph as explained above. The RUS (REP) threshold peaks at $p^{(s)}=0.2\%$ ($0.125\%$) then gradually decreases.
This can be understood by first considering the propagation of spin $X$/$Y$ errors during resource state generation into photon errors, before mapping these photon errors to encoded fusion errors.
A spin $X$/$Y$ error occurred while generating the first $k\leq m$ photons in an encoded qubit will lead to $X/Y$ errors in the subsequently generated $m-k$ photons~\footnote{More precisely, a spin $Y$ error results in a $Y$ error on the first emitted photon and $X$ errors on the next generated photons in the current encoded qubit $i$, as well as a $Z$ error on the first photon of the next qubit $i+1$.}, as well as a $Z$ photon error on the first photon of the next encoded qubit (Sec.~\ref{sec:rsg_spin_error}).
This implies that the probability of flipping the $ZZ$ outcome (hence $\overline{ZZ}$) increases with the number of fusion attempts, since a larger $m$ results in a greater probability of $X/Y$ spin errors within an $m$-photon encoded qubit.
Both error thresholds therefore reduce at higher values of $m$/$N$. Moreover, for REP, majority voting becomes ineffective at correcting erroneous $ZZ$ outcome due to a high weight of $X/Y$ errors from a single spin error.

In practice, the above considered errors will likely be present simultaneously during computation. In Supplementary Note~\ref{supp:FT}, we have considered a combination of photon loss, distinguishability and spin depolarizing errors, and simulated the corresponding fault-tolerant region using RUS fusions.

\section{Discussion}
We have proposed a fusion-based quantum computing architecture tailored to quantum emitters, which is highly resilient to both photon loss and qubit errors, including photon distinguishability between different emitters and decoherence in spin-photon interfaces. The only resource state required in the architecture is a redundantly encoded linear cluster state which can be deterministically generated from a single quantum emitter. The presented high loss threshold of $\ell=8\%$ originates from decomposition of the FFCC lattice in which all fusions on the same layer of the lattice (Fig.~\ref{fig:network}) can be applied simultaneously, thus enabling a repeat-until-success (RUS) strategy. For photon distinguishability between emitters, RUS fusions allows a distinguishability threshold of $4\%$, close to what is already achievable from state-of-the-art quantum emitters ($7\%$)~\cite{Zhai2022}. Encouragingly, since this error only manifests as a photon $X$ error for rotated physical fusions, biased decoders could be used to further improve the threshold~\cite{Tuckett2019}.

Markovian spin noise occurred during resource state generation is modelled by a depolarizing channel with spin error probability $p^{(s)}$ per photon generation time $\tau$. Our architecture can tolerate $p^{(s)}=0.15\%$, which translates to a required spin coherence time of $T_2= 500\tau$, already below the spin coherence times~\footnote{
With $\tau=5\times T_{\text{Mira}}\approx68.9~$ns for quantum-dot entanglement sources~\cite{Meng2024} (determined by pulse-picking every $5$ picosecond pulses generated by the Ti:Sapphire Coherent Mira laser with repetition period $T_{\text{Mira}}=13.77$~ns), $T_2=34.5~\mu\text{s}<113~\mu$s~\cite{Zaporski2023}.} for some quantum emitters used in, i.e., spin-photon entanglement sources with an inherent spin-echo sequence~\cite{Meng2024}. 

Photon loss still remains to be the primary bottleneck for FBQC, with the highest demonstrated outcoupling efficiency currently limited to $71\%$~\cite{Ding2023}. However, recent experimental breakthroughs in photonic hardware~\cite{Alexander2024,Thomas2024} and further architectural optimizations hold promise to bridge this gap. For instance, deterministic entanglement between two emitters is possible by attempting RUS fusions between their respective coupled emitters, followed by entanglement swapping upon fusion success. This would significantly improve the encoded fusion success probability and hence its resistance to photon loss~\cite{Choi2019,Pettersson2025}. A detailed comparison between the aforementioned thresholds and current experimental capabilities is provided in Supplementary Note~\ref{supp:compare}.

Our analysis provides the foundations for capturing realistic errors during resource state generation in emitter-based FBQC architectures. Further study will explore more detailed noise models including hardware-specific error processes, such as imperfect Raman pulses~\cite{Thomas2022}, finite optical cyclicity~\cite{Meng2024} and non-Markovian spin noises relevant to photon emitters~\cite{Stockill2016}.

While architectures with RUS fusions demonstrated greater tolerance to loss and spin errors than that with REP (static repetition code) fusions, they may also be more experimentally demanding. For RUS fusions between two resource state generators, no fiber delay is required, but the failure basis needs to be adaptively switched based on the previous physical fusion outcome. In the REP architecture, fusions on the same layer can in principle be performed synchronously with less resource state generators through interleaving~\cite{bombin_interleaving_2021}. However, this requires additional fiber delays which, in combination with the lower loss threshold ($\ell = 2.3\%$) of the REP architecture, can pose significant challenges.

Despite these experimental hurdles, the synchronous FFCC architecture requires only a fixed number of optical switches per resource state generator, independent of the lattice size. This is in contrast to the original FFCC architecture~\cite{paesani_high-threshold_2022}, which requires an increasing number of switches for fusions between different resource states at larger lattice sizes, as the resource states percolate across multiple unit cells. The practical implementation of these architectures will be a key focus for future research directions. Our work, therefore, creates a clear path for further development of both quantum photonic hardware and fusion-based architectures leveraging deterministic quantum emitters.

\ \\
Note: while writing this manuscript, we became aware of a similar work investigating photonic fault-tolerant architectures with quantum emitters~\cite{QuandelaInPreparation}.

\section*{Acknowledgement}
M.L.C and T.J.B. contributed equally to this work. We gratefully acknowledge financial support from Danmarks Grundforskningsfond (DNRF 139, Hy-Q Center for Hybrid Quantum Networks), the Novo Nordisk Foundation (Challenge
project ”Solid-Q”), and Danmarks Innovationsfond (IFD1003402609,
FTQP). M.L.C acknowledges funding from Danmarks Innovationsfond (Grant No.~4298-00011B). T.J.B. and S.X.C. acknowledges support from UK EPSRC (EP/SO23607/1).  S.P. acknowledges funding from the Marie Skłodowska-Curie Fellowship project QSun (Grant No. 101063763), the VILLUM FONDEN research grants No.~VIL50326 and No.~VIL60743, and support from the NNF Quantum Computing Programme. 

\section*{Data Availability}
The data that support the findings of this article are openly available~\cite{erda}.

%


\setcounter{section}{0}
\setcounter{secnumdepth}{1}
\pagebreak
\clearpage 

\title{Supplementary Notes - \TitleName}

\maketitle
\onecolumngrid
\setcounter{equation}{0}
\setcounter{figure}{0}
\setcounter{table}{0}
\setcounter{page}{1}
\makeatletter
\renewcommand{\theequation}{S\arabic{equation}}
\renewcommand{\thefigure}{S\arabic{figure}}
\renewcommand{\thetable}{S\arabic{table}}
\renewcommand{\bibnumfmt}[1]{[S#1]}
\renewcommand{\citenumfont}[1]{#1}
\renewcommand{\@seccntformat}[1]{%
  \csname the#1\endcsname.\quad
}

{
  \hypersetup{linkcolor=black}
}

\section{Proof for edge and node split rules}
\label{supp:proof}

In this section, we provide proofs for the edge and node split rules using stabilizer formalism. We then show how the node split rule is applied to decompose the FFCC lattice in Fig.~\ref{fig:combined_lattice}a into the synchronous FFCC fusion network in Fig.~\ref{fig:combined_lattice}b.

\begin{figure*}[ht]
    \centering
    \includegraphics[width=0.5\columnwidth]{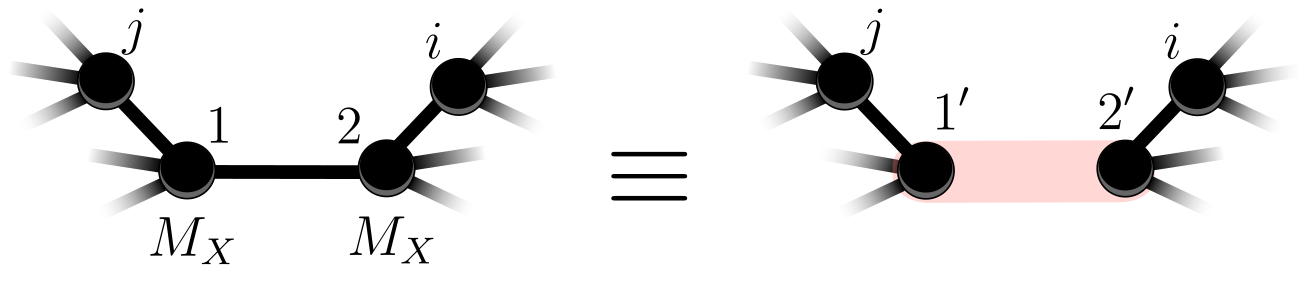}
    \caption{\textbf{Edge split.} $M_X$ is a $X$ measurement performed on the black qubit. Pink bar represents a $\left\{XZ, ZX\right\}$ fusion.}
    \label{fig:SM_edge}
\end{figure*}

\textbf{Edge split rule} is constructed based on the equivalence between a $\left\{XZ, ZX\right\}$ fusion and a $CZ$ gate between the two qubits involved in the fusion, before they are measured in the Pauli-$X$ basis. It can be applied to any two qubits connected with a common edge. In Fig.~\ref{fig:SM_edge}, we consider a general case where each of the qubits involved ($1$ and $2$) is connected to an arbitrary graph.

To show the equivalence of two graphs shown in Fig.~\ref{fig:SM_edge}, we compare their stabilizer generators after measurements. For LHS, the stabilizer generators associated with qubits $1$ and $2$ are:
\begin{align}
    \mathcal{S}_{\text{LHS}}~=&~\langle \stackrel{*}{X_1 Z_{\mathcal{N}_1 \setminus \{2\}} Z_2},~ \{X_j Z_{\mathcal{N}_j \setminus \{1\}}  Z_1 \}_{j\in \mathcal{N}_1 \setminus \{2\}},~X_2 Z_{\mathcal{N}_2 \setminus \{1\}} Z_1, ~\{X_i Z_{\mathcal{N}_i \setminus \{2\}}  Z_2 \}_{i\in \mathcal{N}_2 \setminus \{1\}}\rangle\nonumber\\
    \xrightarrow{X_2}&~ \langle \{X_j Z_{\mathcal{N}_j \setminus \{1\}}  Z_1 \}_{j\in \mathcal{N}_1 \setminus \{2\}},~\stackrel{*}{Z_{\mathcal{N}_2 \setminus \{1\}} Z_1}, ~\{X_1 Z_{\mathcal{N}_1 \setminus \{2\}} X_i Z_{\mathcal{N}_i \setminus \{2\}} \}_{i\in \mathcal{N}_2 \setminus \{1\}}\rangle \otimes \langle X_2 \rangle \nonumber\\
    \xrightarrow{X_1}&~ \langle \{Z_{\mathcal{N}_2 \setminus \{1\}} X_j Z_{\mathcal{N}_j \setminus \{1\}}  \}_{j\in \mathcal{N}_1 \setminus \{2\}}, ~\{Z_{\mathcal{N}_1 \setminus \{2\}} X_i Z_{\mathcal{N}_i \setminus \{2\}} \}_{i\in \mathcal{N}_2 \setminus \{1\}}\rangle \otimes \langle X_2 \rangle \otimes \langle X_1 \rangle,\label{eq:edge_LHS}
\end{align}
where $Z_{\mathcal{N}_i \setminus \{k\}}\equiv \prod_{j\in \mathcal{N}_i  \setminus \{k\}} Z_j$ is the tensor product of Pauli-$Z$ operators on all neighboring qubits $\mathcal{N}_i$ connected to qubit $i$, but excluding the $Z$ operator on qubit $k$. The symbol $(*)$ indicates that the stabilizer generator anti-commutes with measurement in the subsequent step and is thus picked to construct a new list of post-measurement stabilizer generators~\cite{Nielsen2010}. For simplicity we assume that all measurements are deterministic and give positive outcomes. A negative outcome only changes the sign of stabilizers. The curly bracket $\{S_j \}_{j\in \mathcal{N}_i \setminus \{k\}}$ includes all stabilizer generators $S_j$ of qubit $j$ where $j$ belongs to a set of neighbors of qubit $i$ excluding qubit $k$. 

For the graph on RHS, the relevant stabilizer generators evolve as
\begin{align}
    \mathcal{S}_{\text{RHS}}~=&~ \langle \stackrel{*}{X_{1'} Z_{\mathcal{N}_{1'} \setminus \{j\}} Z_{j}},~\{ X_j  Z_{\mathcal{N}_{j} \setminus \{1'\}} Z_{1'}  \}_{j\in \mathcal{N}_{1'}},~X_{2'} Z_{\mathcal{N}_{2'} \setminus \{i\}} Z_{i},~\{ X_i  Z_{\mathcal{N}_{i} \setminus \{2'\}} Z_{2'}  \}_{i\in \mathcal{N}_{2'}}
    \rangle\nonumber\\
    \xrightarrow{Z_{1'} X_{2'}}&~ \langle \{ X_j  Z_{\mathcal{N}_{j} \setminus \{1'\}} Z_{1'}  \}_{j\in \mathcal{N}_{1'}},~\stackrel{*}{X_{2'} Z_{\mathcal{N}_{2'} \setminus \{i\}} Z_{i}},~\{ X_{1'} Z_{2'} Z_{\mathcal{N}_{1'} \setminus \{j\}} Z_{j} X_i  Z_{\mathcal{N}_{i} \setminus \{2'\}}  \}_{i\in \mathcal{N}_{2'}}
    \rangle \otimes \langle Z_{1'} X_{2'} \rangle \nonumber\\
    \xrightarrow{X_{1'} Z_{2'}}&~ \langle \{ Z_{\mathcal{N}_{2'} \setminus \{i\}} Z_{i} X_j  Z_{\mathcal{N}_{j} \setminus \{1'\}}   \}_{j\in \mathcal{N}_{1'} },~\{ Z_{\mathcal{N}_{1'} \setminus \{j\}} Z_{j} X_i  Z_{\mathcal{N}_{i} \setminus \{2'\}}  \}_{i\in \mathcal{N}_{2'}}
    \rangle \otimes \langle Z_{1'} X_{2'}\rangle \otimes \langle X_{1'} Z_{2'} \rangle\nonumber\\
    =& ~\langle \{Z_{\mathcal{N}_2 \setminus \{1\}} X_j Z_{\mathcal{N}_j \setminus \{1\}}  \}_{j\in \mathcal{N}_1 \setminus \{2\}}, ~\{Z_{\mathcal{N}_1 \setminus \{2\}} X_i Z_{\mathcal{N}_i \setminus \{2\}} \}_{i\in \mathcal{N}_2 \setminus \{1\}}\rangle \otimes \langle Z_{1'} X_{2'}\rangle \otimes \langle X_{1'} Z_{2'} \rangle,
\end{align}
which gives identical stabilizer generators as in Eq.~(\ref{eq:edge_LHS}). The equality in the last step results from the fact that the set of neighboring qubits $\mathcal{N}_{1'}$ ($\mathcal{N}_{2'}$) is equivalent to the set $\mathcal{N}_{1} \setminus \{2\}$ ($\mathcal{N}_{2} \setminus \{1\}$). $X_{1'} Z_{2'}$ and $Z_{1'} X_{2'}$ are measurement operators for a successful $\{XZ,ZX\}$ physical fusion on qubits $1'$ and $2'$.

\begin{figure*}[ht]
    \centering
    \includegraphics[width=0.45\columnwidth]{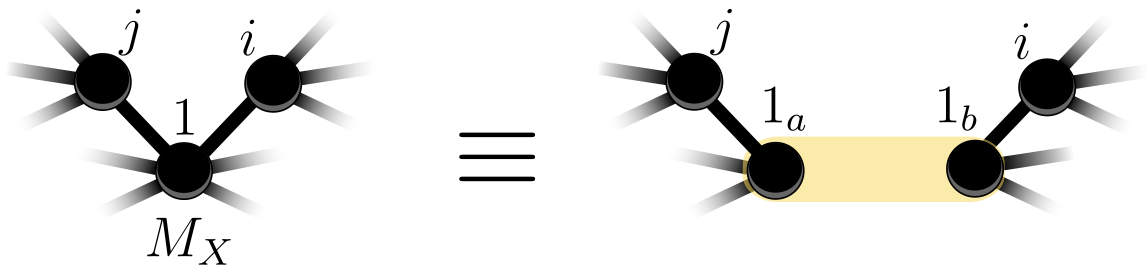}
    \caption{\textbf{Node split.} Qubits $i$ and $j$ are included to show equivalence in the surviving stabilizer group between both graphs.}
    \label{fig:SM_node}
\end{figure*}

\textbf{Node split rule} states that measuring a qubit of a cluster state in the Pauli-$X$ basis is equivalent to applying a $\left\{XX, ZZ\right\}$ fusion (yellow bar in Fig.~\ref{fig:SM_node}) between two additional qubits introduced to the same node of the graph.

The graph on LHS is stabilized by generators which change after measuring qubit $1$ in the $X$ basis:
\begin{align}
    \mathcal{S}_{\text{LHS}} =&~ \langle X_1 Z_{\mathcal{N}_1},~\stackrel{*}{X_j Z_{\mathcal{N}_j \setminus \{1\}} Z_1},~X_i Z_{\mathcal{N}_i \setminus \{1\}} Z_1,~ \{ X_k Z_{\mathcal{N}_k} Z_1 \}_{k \in \mathcal{N}_1 \setminus \{j,i\}}  \rangle\nonumber\\
    \xrightarrow{X_1}&~ \langle Z_{\mathcal{N}_1},~X_j Z_{\mathcal{N}_j \setminus \{1\}} X_i Z_{\mathcal{N}_i \setminus \{1\}},~ \{ X_j Z_{\mathcal{N}_j \setminus \{1\}} X_k Z_{\mathcal{N}_k} \}_{k \in \mathcal{N}_1 \setminus \{j,i\}}  \rangle \otimes \langle X_1 \rangle.
\end{align}
Similarly, for RHS,
\begin{align}
    \mathcal{S}_{\text{RHS}} =&~ \langle \stackrel{*}{X_{1_a} Z_{\mathcal{N}_{1_a}  \setminus \{j\}} Z_j},~ X_{j} Z_{\mathcal{N}_{j} \setminus \{1_a\}} Z_{1_a},~ \{ X_k Z_{\mathcal{N}_k \setminus \{1_a\}} Z_{1_a} \}_{k \in \mathcal{N}_{1_a} \setminus \{j\}},\quad\nonumber\\
    &\quad\quad ~ X_{1_b} Z_{\mathcal{N}_{1_b}  \setminus \{i\}} Z_i,~X_{i} Z_{\mathcal{N}_{i} \setminus \{1_b\}} Z_{1_b},~ \{ X_k Z_{\mathcal{N}_k \setminus \{1_b\}} Z_{1_b} \}_{k \in \mathcal{N}_{1_b} \setminus \{i\}}  \rangle\nonumber\\
    \xrightarrow{Z_{1_a}Z_{1_b}}&~\langle \stackrel{*}{ X_{j} Z_{\mathcal{N}_{j} \setminus \{1_a\}} Z_{1_a}},~ \{ X_k Z_{\mathcal{N}_k \setminus \{1_a\}} Z_{1_a} \}_{k \in \mathcal{N}_{1_a} \setminus \{j\}}, ~X_{1_a} X_{1_b} Z_{\mathcal{N}_{1_a}  \setminus \{j\}} Z_j Z_{\mathcal{N}_{1_b}  \setminus \{i\}} Z_i,\nonumber\\
    &\quad\quad~X_{i} Z_{\mathcal{N}_{i} \setminus \{1_b\}} Z_{1_b},~ \{ X_k Z_{\mathcal{N}_k \setminus \{1_b\}} Z_{1_b} \}_{k \in \mathcal{N}_{1_b} \setminus \{i\}} \rangle \otimes \langle Z_{1_a}Z_{1_b}\rangle\nonumber\\
    \xrightarrow{X_{1_a}X_{1_b}}&~ \langle  \{ X_{j} Z_{\mathcal{N}_{j} \setminus \{1_a\}} X_k Z_{\mathcal{N}_k \setminus \{1_a\}} \}_{k \in \mathcal{N}_{1_a} \setminus \{j\}}, ~ Z_{\mathcal{N}_{1_a}  \setminus \{j\}} Z_j Z_{\mathcal{N}_{1_b}  \setminus \{i\}} Z_i,\nonumber\\
    &\quad\quad~ X_{j} Z_{\mathcal{N}_{j} \setminus \{1_a\}}  X_{i} Z_{\mathcal{N}_{i} \setminus \{1_b\}} ,~ \{ X_{j} Z_{\mathcal{N}_{j} \setminus \{1_a\}} X_k Z_{\mathcal{N}_k \setminus \{1_b\}} \}_{k \in \mathcal{N}_{1_b} \setminus \{i\}} \rangle \otimes \langle Z_{1_a}Z_{1_b}\rangle\otimes \langle X_{1_a}X_{1_b}\rangle\nonumber\\
    =&~\langle Z_{\mathcal{N}_1},~X_j Z_{\mathcal{N}_j \setminus \{1\}} X_i Z_{\mathcal{N}_i \setminus \{1\}},~ \{ X_j Z_{\mathcal{N}_j \setminus \{1\}} X_k Z_{\mathcal{N}_k} \}_{k \in \mathcal{N}_1 \setminus \{j,i\}}  \rangle \otimes \langle Z_{1_a}Z_{1_b}\rangle\otimes \langle X_{1_a}X_{1_b}\rangle.
\end{align}
The equality in the last step originates from (1) $Z_{\mathcal{N}_1}=Z_{\mathcal{N}_{1_a}}Z_{\mathcal{N}_{1_b}}$, and (2) the set of qubits in $\mathcal{N}_{j} \setminus \{1\} $ ($\mathcal{N}_{i} \setminus \{1\} $) of LHS is identical to the set $\mathcal{N}_{j} \setminus \{1_a\}$ ($\mathcal{N}_{i} \setminus \{1_b\}$).
\subsection{Decomposition into synchronous FFCC fusion network using node splits and $X$ measurements}
The decomposition can be understood in two steps: First, apply node split to every check qubit (white circle) of the lattice in Fig.~\ref{fig:combined_lattice}a. Next, $X$ measurements are performed on two neighboring qubits previously connected to the check qubit.

\begin{figure*}[ht]
    \centering
    \includegraphics[width=0.8\columnwidth]{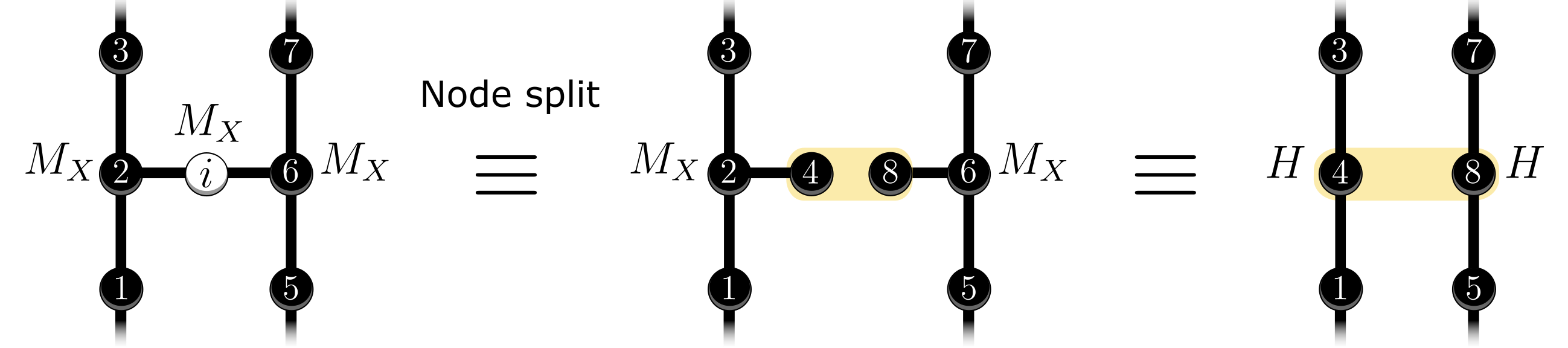}
    \caption{\textbf{Graph decomposition using node split and $X$ measurements.} Here we consider the decomposition of a section of the FFCC lattice into a fusion between linear cluster states. First, the check qubit (white) is split into qubits $4$ and $8$ which will undergo a $\{XX,ZZ\}$ fusion. Second, $X$ measurements on qubits $2$ and $6$ remove them from the graph and connect their neighbors, with a Hadamard gate on each of the neighbors.}
    \label{fig:SM_decompose}
\end{figure*}

In Fig.~\ref{fig:SM_decompose}a we visualize the decomposition process in the graph state picture. We begin with a ladder-like graph in the FFCC lattice. As described in the main text, in the measurement-based model, all qubits including data (black) and check qubits (white) are measured in the Pauli-$X$ basis to propagate information from bottom layers to the top. Since we are interested in the graph transformation in the vicinity of the check qubit $i$, only $X$ measurements on its neighbors (qubits $2$ and $6$) which are relevant to the transformation are shown in the first graph of Fig.~\ref{fig:SM_decompose}. The idea of decomposition is to split the ladder-like graph into a fusion between two simpler states, i.e., linear cluster states. Therefore, we choose to split qubit $i$ into two additional qubits, instead of its neighbors. Using the node split rule in Fig.~\ref{fig:SM_node}, we obtain the second graph with additional qubits $4$ and $8$, which is equivalent to a fusion between two branched chain states followed by $X$ measurements on qubits $2$ and $6$.

Next, instead of applying node split again, these $X$ measurements are performed to transform the graph into two linear cluster states, with Hadamard gates on both qubits $4$ and $8$, as shown in the third graph of Fig.~\ref{fig:SM_decompose}. These Hadamard gates do not change the fusion measurement outcomes and only alter the fusion failure basis. The graph transformation under a $X$ measurement can be verified similarly. After measuring qubit $2$ in $X$ in the second graph of Fig.~\ref{fig:SM_decompose}, the stabilizer generators of the left branch chain state become
\begin{align}
    \mathcal{S} =&~ \langle X_1 Z_2 Z_{\mathcal{N}_1 \setminus \{2\}},~ X_2 Z_1 Z_3 Z_4, ~X_3 Z_2 Z_{\mathcal{N}_3 \setminus \{2\}},~ \stackrel{*}{X_4 Z_2} \rangle\nonumber\\
    \xrightarrow{X_2}&~ \langle X_1 X_4 Z_{\mathcal{N}_1 \setminus \{2\}},~  Z_1 Z_3 Z_4, ~X_3 X_4 Z_{\mathcal{N}_3 \setminus \{2\}}\rangle\nonumber\\
    =&~H_4 \bigg[ \langle X_1 Z_4 Z_{\mathcal{N}_1 \setminus \{2\}},~  Z_1 Z_3 X_4, ~X_3 Z_4 Z_{\mathcal{N}_3 \setminus \{2\}}\rangle\bigg],
\end{align}
which is indeed identical to a linear cluster state with a Hadamard rotation on qubit $4$.

Therefore, repeating the above steps on all other check qubits in Fig.~\ref{fig:combined_lattice}a, we obtain the synchronous FFCC fusion network depicted in Fig.~\ref{fig:combined_lattice}b. It is important to emphasize that during resource state generation, these $X$ measurements are not implemented, i.e., they are effectively virtual measurements that always give $+1$ outcome. As a result of the equivalence depicted in Fig.~\ref{fig:SM_decompose}, it is only necessary to generate linear cluster states and perform fusion measurements. In other words, the synchronous FFCC fusion network is equivalent to the FFCC lattice in the measurement-based model, where these $X$ measurements are always successful resulting in the $+1$ outcome.

As a final check, the equivalence between the first and third graphs in Fig.~\ref{fig:SM_decompose} is verified by comparing their surviving stabilizer groups:
\begin{align}
    \mathcal{S}_{\text{LHS}} =&~ \langle X_1 Z_{\mathcal{N}_1 \setminus \{2\}} Z_2,~\stackrel{*}{X_2 Z_1 Z_i Z_3},~X_3 Z_{\mathcal{N}_3 \setminus \{2\}} Z_2,~X_i Z_2 Z_6,~X_6 Z_5 Z_i Z_7 ,~X_5 Z_{\mathcal{N}_5 \setminus \{6\}} Z_6,~X_7 Z_{\mathcal{N}_7 \setminus \{6\}} Z_6 \rangle\nonumber\\
    \xrightarrow{X_i}&~ \langle \stackrel{*}{X_1 Z_{\mathcal{N}_1 \setminus \{2\}} Z_2},~X_3 Z_{\mathcal{N}_3 \setminus \{2\}} Z_2,~Z_2 Z_6,~X_2 Z_1 Z_3 X_6 Z_5 Z_7,~X_5 Z_{\mathcal{N}_5 \setminus \{6\}} Z_6,~X_7 Z_{\mathcal{N}_7 \setminus \{6\}} Z_6 \rangle\otimes \langle X_i \rangle\nonumber\\
     \xrightarrow{X_2}&~ \langle X_1 Z_{\mathcal{N}_1 \setminus \{2\}} X_3 Z_{\mathcal{N}_3 \setminus \{2\}},~\stackrel{*}{X_1 Z_{\mathcal{N}_1 \setminus \{2\}} Z_6},~ Z_1 Z_3 X_6 Z_5 Z_7,~X_5 Z_{\mathcal{N}_5 \setminus \{6\}} Z_6,~X_7 Z_{\mathcal{N}_7 \setminus \{6\}} Z_6 \rangle\otimes \langle X_i \rangle\otimes \langle X_2 \rangle\nonumber\\
     \xrightarrow{X_6}&~ \langle X_1 Z_{\mathcal{N}_1 \setminus \{2\}} X_3 Z_{\mathcal{N}_3 \setminus \{2\}},~ Z_1 Z_3 Z_5 Z_7,~X_1 Z_{\mathcal{N}_1 \setminus \{2\}} X_5 Z_{\mathcal{N}_5 \setminus \{6\}},~X_1 Z_{\mathcal{N}_1 \setminus \{2\}} X_7 Z_{\mathcal{N}_7 \setminus \{6\}} \rangle\otimes \langle X_i \rangle\otimes \langle X_2 \rangle\otimes \langle X_6 \rangle.\\
     \mathcal{S}_{\text{RHS}} =&~ \langle \stackrel{*}{X_1 Z_{\mathcal{N}_1 \setminus \{4\}} Z_4}, ~X_4 Z_1 Z_3, ~X_3 Z_{\mathcal{N}_3 \setminus \{4\}} Z_4,~X_5 Z_{\mathcal{N}_5 \setminus \{8\}} Z_8, ~X_8 Z_5 Z_7, ~X_7 Z_{\mathcal{N}_7 \setminus \{8\}} Z_8 \rangle\nonumber\\
     \xrightarrow{X_4 X_8}&~ \langle  \stackrel{*}{X_4 Z_1 Z_3}, ~X_1 Z_{\mathcal{N}_1 \setminus \{4\}} X_3 Z_{\mathcal{N}_3 \setminus \{4\}} ,~X_1 Z_{\mathcal{N}_1 \setminus \{4\}} X_5 Z_{\mathcal{N}_5 \setminus \{8\}} Z_4 Z_8, ~X_8 Z_5 Z_7, ~X_1 Z_{\mathcal{N}_1 \setminus \{4\}} X_7 Z_{\mathcal{N}_7 \setminus \{8\}} Z_4 Z_8 \rangle\otimes \langle X_4 X_8 \rangle\nonumber\\
     \xrightarrow{Z_4 Z_8}&~ \langle X_1 Z_{\mathcal{N}_1 \setminus \{4\}} X_3 Z_{\mathcal{N}_3 \setminus \{4\}} ,~X_1 Z_{\mathcal{N}_1 \setminus \{4\}} X_5 Z_{\mathcal{N}_5 \setminus \{8\}} , ~Z_1 Z_3 Z_5 Z_7, ~X_1 Z_{\mathcal{N}_1 \setminus \{4\}} X_7 Z_{\mathcal{N}_7 \setminus \{8\}} \rangle \otimes \langle X_4 X_8 \rangle\otimes \langle Z_4 Z_8 \rangle,
\end{align}
where $Z_{\mathcal{N}_1 \setminus \{2\}}$ ($Z_{\mathcal{N}_7 \setminus \{6\}}$) in the first graph is identical to $Z_{\mathcal{N}_1 \setminus \{4\}}$ ($Z_{\mathcal{N}_7 \setminus \{8\}}$) in the final graph.

\newpage
\section{Erasure and error rates of encoded fusions}
\label{supp:RUS}
Here we calculate the probability of obtaining or recovering measurement outcomes $\overline{XX}$ and $\overline{ZZ}$ in an encoded fusion, using REP and RUS in the presence of photon loss. The encoded fusion error rates are also computed, with $X$ and $Z$ errors applied to each physical fusion.
In general, with an end-to-end efficiency of $\eta$ and a fusion failure rate of $P_{\text{fail}}$, for a physical fusion there are three possible events:
\begin{equation}
    \text{Recovered outcome} = \begin{cases}
        XX \text{ and } ZZ, & P_{\text{s}} = \eta^{2}(1 - P_{\text{fail}}); \\
        XX \text{ only}, & P_{\text{f}} = \eta^{2}P_{\text{fail}}; \\
        \text{erasure,} & P_l = 1 - \eta^{2},
    \end{cases}
\end{equation}
where the subscripts s, f and $l$ correspond to fusion success, failure and erasure due to photon loss, with respectively. As discussed in Sec.~\ref{sec:repcodes}, it is beneficial to do rotated physical fusions where $XX$ outcome is retrieved upon fusion failure.

For \textbf{REP encoded fusions}, as outlined in the main text, $m$ fusions are attempted ballistically between pairs of photons in the encoded qubits.
All $m$ physical $XX$ outcomes are required to obtain the $\overline{XX}$ outcome, whereas only a single physical $ZZ$ outcome is required for $\overline{ZZ}$.
The corresponding recovery probabilities are then
\begin{equation}
    R_{\overline{XX}} = \left(P_{\text{s}} + P_{\text{f}}\right)^{m}; \qquad R_{\overline{ZZ}} = 1 - \left(1 - P_{\text{s}}\right)^{m}.
\end{equation}
The logical error rates of REP fusion can be found by counting the number of physical fusion errors and determining whether this gives a logical error.
Let us assume that $X$ and $Z$ errors occur on each fusion outcome with probabilities $p_X$ and $p_Z$, which flip $ZZ$ and $XX$ outcomes, respectively. If the probability that a photon acquiring $Z$ error prior to fusion is $e_Z$, then the probability of having a flipped $XX$ fusion outcome is the probability of odd numbers of photon $Z$ error which is $p_Z=2 e_Z$.
Similarly on the logical level, since the $Z$ repetition code offers no assistance to the recovery of $\overline{XX}$ outcome, the probability of a logical error in $\overline{XX}$ is the probability of having an odd number of $Z$ fusion errors in $m$ physical fusion attempts:
\begin{equation}
    \mathcal{E}_{\overline{XX}} = \sum_{k \text{ odd}}^{m} \binom{m}{k}(p_Z)^{k}(1-p_Z)^{m-k}.
\end{equation}
With majority voting, flips in the $\overline{ZZ}$ outcome occur when half or more of the $t$ recovered physical $ZZ$ outcomes have an error.
\begin{equation}
    \mathcal{E}_{\overline{ZZ}} = \sum_{t=1}^{m}\binom{m}{t}(P_{\text{s}})^{t}(1-P_{\text{s}})^{m-t} \sum_{k=\lfloor\frac{t+1}{2}\rfloor}^{t}\binom{t}{k} (p_X)^{k}(1-p_X)^{t-k} \frac{1}{R_{\overline{ZZ}}}.
\end{equation}
These expressions denote the conditional probability of an error given that $\overline{XX}$ and $\overline{ZZ}$ were recovered. 
We can improve the error tolerance by treating a $\overline{ZZ}$ outcome as erased when there are equal numbers of good and erroneous physical $ZZ$ outcomes, as we have no confidence in the value of the $\overline{ZZ}$ operator.

\textbf{RUS encoded fusions} have the same requirements for obtaining encoded fusion outcomes, but now the code size can vary. Moreover, only a single $ZZ$ operator is required to recover $\overline{ZZ}$ since RUS terminates further fusion attempts once $ZZ$ is obtained.
Four distinct events can occur after a logical RUS fusion: \textcircled{1} $0\leq j<N$ failures occur before a successful fusion, recovering both $\overline{XX}$ and $\overline{ZZ}$.
\textcircled{2}: $N$ consecutive failures occur, recovering only $\overline{XX}$.
\textcircled{3}: $0\leq j\leq N-2$ consecutive failures occur, followed by a loss in the $(j+1)$-th fusion. The physical fusion failure basis is changed to $ZZ$. There are then $0\leq k < N - j - 2$ additional losses before a single non-lossy fusion that recovers $\overline{ZZ}$ only.
\textcircled{4}: As for \textcircled{3}, but all fusions after the $j$ failures have loss. No outcomes are recovered, corresponding to logical erasure.
These events occur with probabilities
\begin{equation}
\begin{split}
        \overline{XX} \text{ and } \overline{ZZ}: \quad P_{1} &= P_{\text{s}}\sum_{j = 0}^{N-1}(P_{\text{f}})^{j} ;
        \\\overline{XX} \text{ only}: \quad P_{2} &= (P_{\text{f}})^{N};  \\  \overline{ZZ} \text{ only}: \quad
        P_3 &= \sum_{j=0}^{N-2}(P_{\text{f}})^{j}\sum_{k=1}^{N-j-1}(P_l)^{k}(1-P_l); \quad \\ \text{Erasure}: \quad P_4 &= \sum_{j=0}^{N-1}(P_{\text{f}})^{j}(P_l)^{N-j},
\end{split}
\label{eq:prob}
\end{equation}
where $P_1+P_2+P_3+P_4=1$. The recovery probabilities for the respective outcomes are then $R_{\overline{XX}} = P_1 + P_2$, and $R_{\overline{ZZ}} = P_1 + P_3$.
In the simulation, we draw samples of RUS encoded fusion outcomes from the weighted probability distribution $\{P_1,P_2,P_3,P_4\}$ using Eq.~(\ref{eq:prob}).

The encoded fusion error rates are calculated similarly to before. However, since $\overline{ZZ}$ is always recovered by one $ZZ$ outcome, this means that majority voting is no longer possible, thus the encoded fusion error rate is equal to the physical fusion error rate.
\begin{equation}
    \mathcal{E}_{\overline{XX}}  = \left[\sum_{j = 0}^{N-1}(P_{\text{f}})^{j}P_{\text{s}}  \sum_{k \text{ odd}}^{j+1} \binom{j+1}{k}(p_Z)^{k}(1-p_Z)^{j-1-k} + (P_{\text{f}})^{N} \sum_{k \text{ odd}}^{N} \binom{N}{k}(p_Z)^{k}(1-p_Z)^{N-k} \right]  \frac{1}{ R_{\overline{XX}}};\quad
    \mathcal{E}_{\overline{ZZ}} = p_X.
\end{equation}

\begin{figure*}[ht]
    \centering
    \includegraphics[width=\columnwidth]{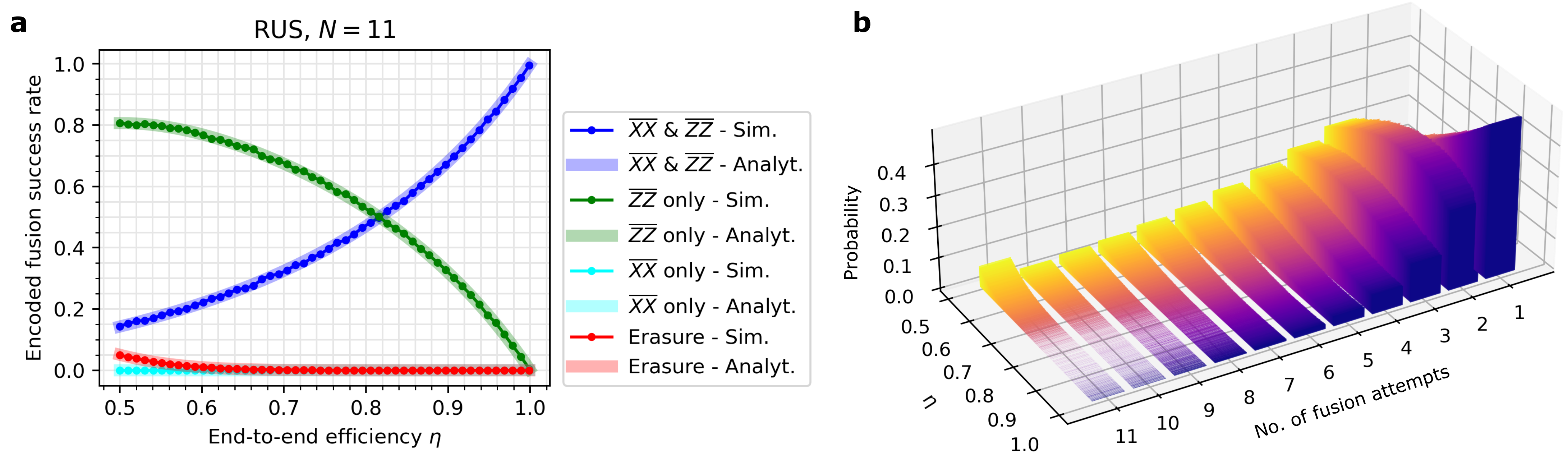}
    \caption{\textbf{Monte-Carlo simulation of RUS fusion.} \textbf{(a)} Simulated encoded fusion success probability as a function of end-to-end efficiency $\eta$. The simulation uses $N=11$ with $10^4$ trials. Analytical curves are obtained using Eq.~(\ref{eq:prob}). \textbf{(b)} Distribution of number of fusion attempts $m$ in a RUS encoded fusion. $z$-axis is the probability that the RUS fusion stops at $m$-th attempt. At $20\%$ loss, there is a $84\%$ chance that the RUS fusion terminates within $m=3$ attempts.}
    \label{fig:RUS_bar}
\end{figure*}

These expressions are verified by Monte-Carlo simulations (Fig.~\ref{fig:RUS_bar}a), which we then use to inspect the distribution of number of physical fusion attempts when $N=11$, as plotted in Fig.~\ref{fig:RUS_bar}b.
In the absence of loss, there is a $87.5\%$ chance that RUS fusion terminates within $m=3$ attempts.

\newpage
\section{RUS fusion with re-initialization of emitters}
\label{supp:RUS_inf}
\begin{figure*}[ht]
    \centering
    \includegraphics[width=0.6\columnwidth]{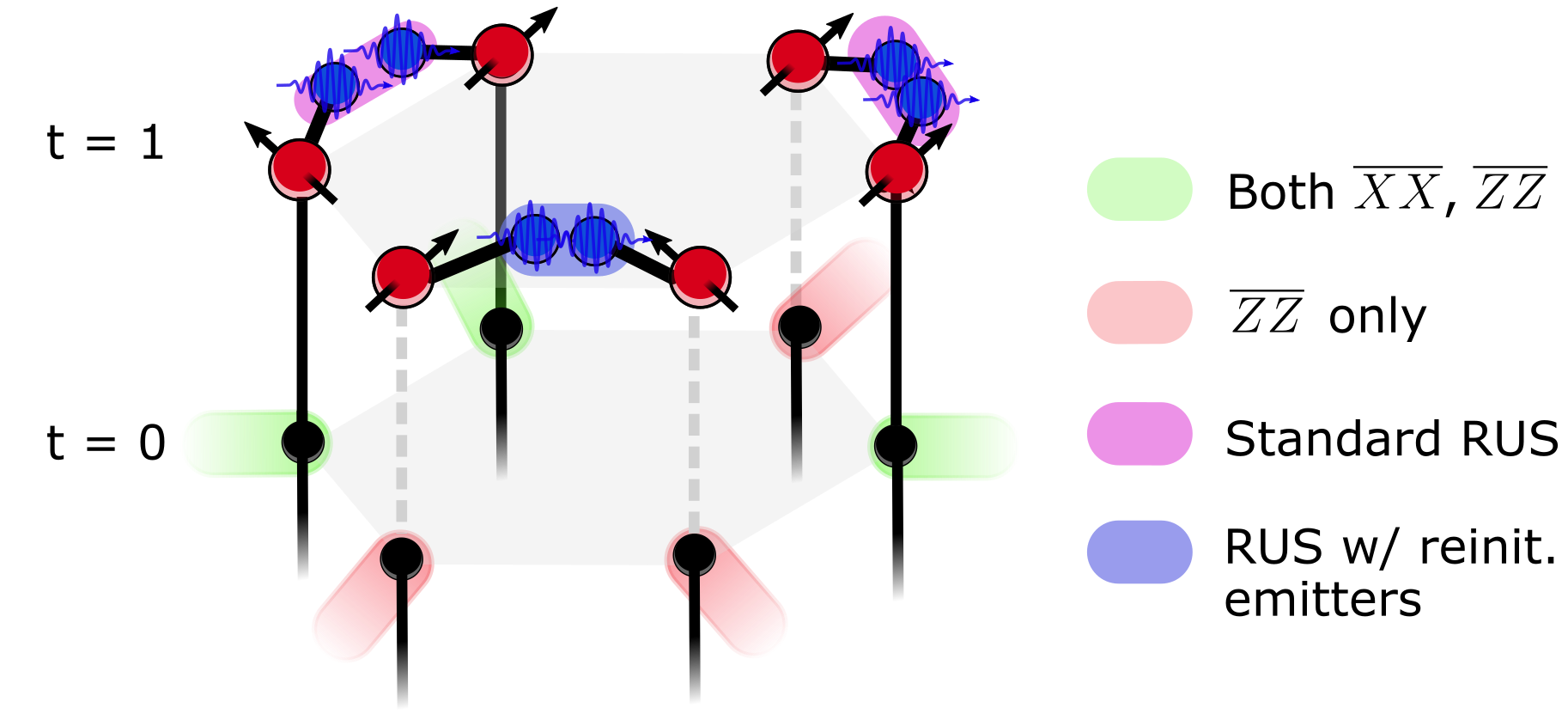}
    \caption{\textbf{Schematics of RUS fusion with re-attempts.} A RUS fusion (blue) at time $t=1$ can be re-attempted when the two emitters are previously disconnected from the graph due to two encoded fusions (pink) resulting in biased $\overline{ZZ}$ outcome at $t=0$.}
    \label{fig:reinit}
\end{figure*}
For RUS encoded fusions, when a photon is lost during a physical fusion, the $XX$ outcome is irreversibly lost thus $\overline{XX}$ cannot be recovered.
On the other hand, the $\overline{ZZ}$ outcome can still be obtained by retrieving $ZZ$ in the next fusion attempt (Sec.~\ref{subsec:RUS_f}), due to re-emission of the photons followed by single-qubit measurement in the Pauli-$Z$ basis.
Crucially, when $ZZ$ outcomes are retrieved from two neighboring encoded fusions, the two emitter spins involved are indirectly measured in the Pauli-$Z$ basis, since the measured photon and its entangled emitter share the stabilizer $ZZ$. This implies that once the encoded fusion results in $\overline{ZZ}$ outcome only, the emitter spins become disconnected from the original graph with their states effectively re-initialized.

In the case of two neighboring encoded fusions being biased in $\overline{ZZ}$ at an initial time $t=0$ (Fig.~\ref{fig:reinit}), both emitters at $t=1$ are re-initialized and detached from any previous correlations,
allowing the next encoded fusion to attempt RUS as many times as possible. The probability of recovering both $\overline{XX}$ and $\overline{ZZ}$ is therefore drastically increased. For realistic simulation, this number is set to be equal to the number of maximum physical fusion attempts $N$.  In the absence of photon loss, the probability of recovering both encoded fusion outcomes scales as $1-2^{-N}$ thus making the encoded fusion near deterministic.
Interestingly, in a realistic experiment this trick can be applied to the first layer of encoded fusions in the FFCC lattice to make them near deterministic, since all emitters are initially disconnected. Here for simulations we are only interested in the bulk of the lattice thus it is applied only to the subsequent layers. It is important to note that this scheme relies on the previous two neighboring encoded fusions only retrieving the $\overline{ZZ}$, which is
a second-order effect, and thus, this additional trick gives a modest improvement of the loss threshold to 8\%. 
\newpage
\section{Fault-tolerant region with multiple errors}\label{supp:FT}
A practical quantum computer will likely be simultaneously subject to multiple error channels, it is therefore useful to know whether a given set of errors lies within the fault-tolerant region of the quantum computing architecture. 

\begin{figure*}[ht]
    \centering
    \includegraphics[width=0.55\columnwidth]{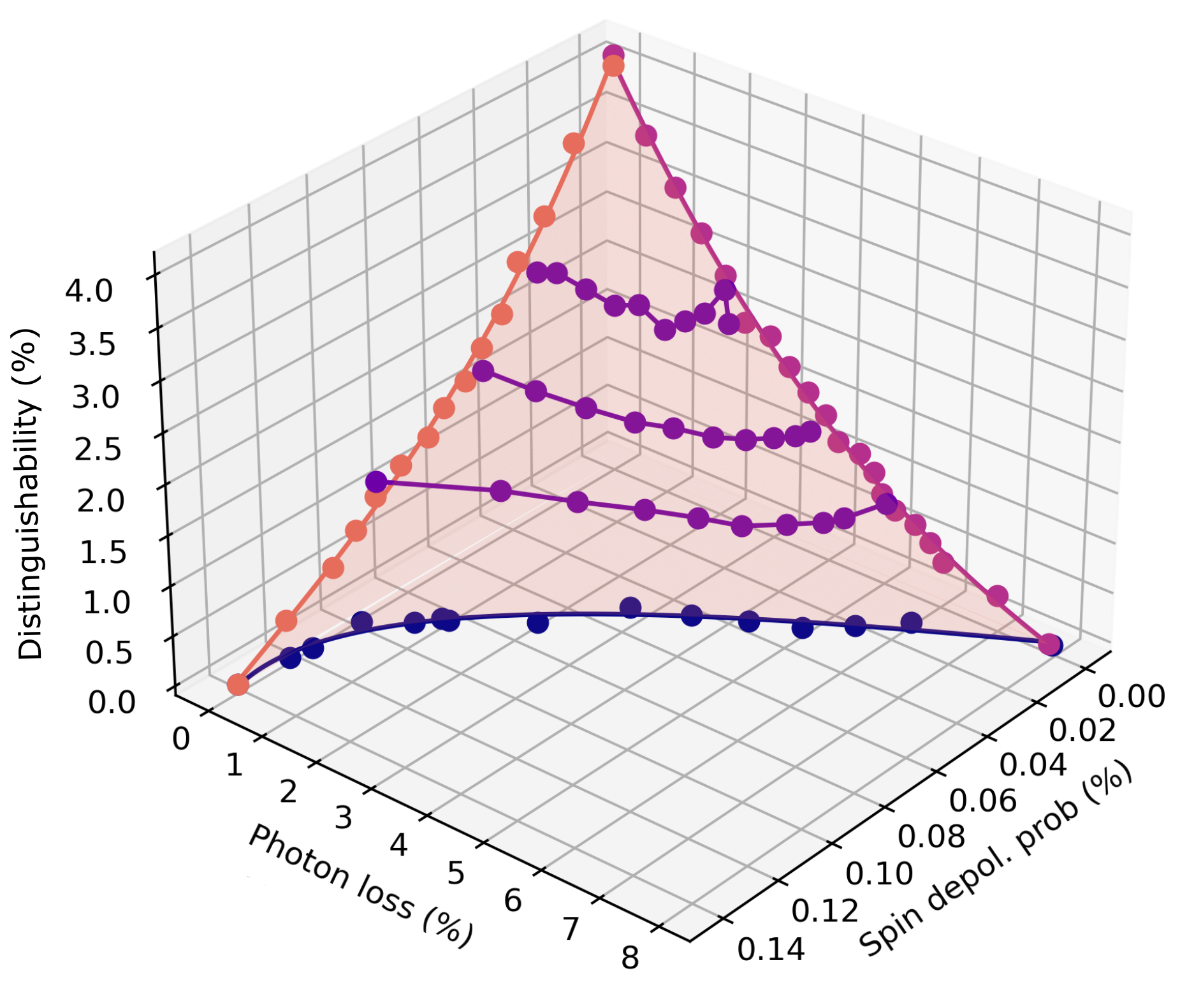}
    \caption{\textbf{Fault-tolerant region of the sFFCC architecture, using RUS fusions with emitter re-initialization.} We consider photon loss, distinguishability between emitters, and spin depolarizing error. The number of maximum fusion attempts and the number of RUS re-attempts are both set to $N=10$.}
    \label{fig:3DFT}
\end{figure*}

In Fig.~\ref{fig:3DFT}, we simulate the combined effect of photon loss, distinguishability between emitters, and spin depolarizing errors, and present the region (shaded in pink) in which the logical error can be corrected. The colored curves at the boundaries are fault-tolerant thresholds obtained with either two of the above errors. Here we consider the synchronous FFCC architecture, using RUS encoded fusions in which the re-initialization of emitters is allowed for maximum loss threshold. Both the numbers of maximum physical fusion attempts and RUS re-attempts are set to $N=10$ since the thresholds shown in the main text begin to saturate at $N\geq10$.

The noise simulation is a two-step process, similar to the one with only spin noise discussed in Sec.~\ref{sec:thresholds}, which begins by first sampling encoded fusion outcomes under the presence of both photon loss and distinguishability from the weighted probability distribution $\{P^{\bar{\text{X}}\bar{\text{Z}}}_{N},P^{\bar{\text{X}}}_{N},P^{\bar{\text{Z}}}_{N},P^{\bar{\text{e}}}_{N}\}$, where the analytical formula for each probability is:
\begin{align}
    \overline{XX} \text{ and } \overline{ZZ}: \quad\quad P^{\bar{\text{X}}\bar{\text{Z}}}_{N} &= P_{\text{s}} + P_{\text{s}} \sum^{N-1}_{i=1} (P_{\text{f}})^{i}
    ;\nonumber\\
    \overline{XX} \text{ only }: \quad\quad P^{\bar{\text{X}}}_{N} &= (P_{\text{f}})^{N} = (\eta^2 P^{\text{e}}_{Z} P_{X})^N;\nonumber\\
    \text{Erasure }: \quad\quad 
    P^{\bar{\text{e}}}_{N} &= (P_l + P^{\text{e}}_{\text{f}}) (P_l)^{N-1} +  (P_l+P^{\text{e}}_{\text{f}})\sum^{N-1}_{i=1} (P_{\text{f}})^{i} (P_l)^{N-1-i}
    ;\nonumber\\
    \overline{ZZ} \text{ only }: \quad\quad 
    P^{\bar{\text{Z}}}_{N} &= 1-P^{\bar{\text{X}}}_{N}-P^{\bar{\text{e}}}_{N}-P^{\bar{\text{X}}\bar{\text{Z}}}_{N}.
\end{align}
$P^j_{N}$ is the probability that an encoded fusion results in measurement outcome $j$ within $N$ physical fusion attempts. With finite photon distinguishability $1-V\neq0$, a physical fusion could in principle result in erasure of both outcomes, i.e., when a $\{ZZ,XX\}$ fusion fails without recovering $XX$. In such a case, the $ZZ$/$XX$ outcome is erased with probabilities $P^\text{e}_Z = \frac{1}{2}=1-P_Z$ and $P^{\text{e}}_X = \frac{1-V}{4}=1-P_X$. Including photon loss, there are in total five possible fusion outcomes: (1) successful fusion without erasure in $XX$, denoted by $P_\text{s}$, (2) successful fusion with $XX$ erasure $P^{\text{e}}_{\text{s}}$, (3) failed fusion without $XX$ erasure $P_\text{f}$, (4) failed fusion with $XX$ erasure $P^{\text{e}}_{\text{f}}$, and (5) fusion erasure $P_l$ owing to photon loss. For the first four cases, both photons are not lost. As a result, the probability for each of the above five cases is:
\begin{align}
    P_{\text{s}} &=  \eta^2 P_{Z} P_{X}; \quad\quad
    P^{\text{e}}_{\text{s}} = \eta^2 P_{Z} P^{\text{e}}_{X};\quad\quad
    P_{\text{f}} = \eta^2 P^{\text{e}}_{Z} P_{X};\quad\quad
    P^{\text{e}}_{\text{f}} = \eta^2 P^{\text{e}}_{Z} P^{\text{e}}_{X};\quad\quad
    P_l = 1-\eta^2.
\end{align}
After sampling a list of outcomes for all encoded fusions, a search is performed over each layer of the lattice to locate encoded fusions that end up only in $\overline{ZZ}$. If the condition for RUS emitter re-initialization is fulfilled (Supplementary Note~\ref{supp:RUS_inf}): two neighboring encoded fusions are biased in $\overline{ZZ}$ at layer $i$, and the encoded fusion at layer $i+1$ only gives either one or no measurement outcome, then its outcome is re-sampled for $N$ additional RUS re-attempts.

Once a final list of encoded fusion outcomes is obtained, a fusion error sampling process for distinguishability and spin depolarizing noise begins for each encoded fusion:
\begin{enumerate}
    \item If RUS fusion results in both $\overline{XX}$ and $\overline{ZZ}$, the $\overline{ZZ}=ZZ$ outcome has a probability of $\frac{1-V}{2}$ being flipped due to photon distinguishability. This $ZZ$ outcome would have resulted from the last physical fusion attempt $m\leq N$, since encoded fusion stops once $\overline{ZZ}$ is recovered. $m$ is sampled from a list of $N$ probabilities $\mathcal{P}_{\bar{X}\bar{Z}}=\{P_i\}_{i\in [1,N]}$, where $P_i = P^{\bar{\text{X}}\bar{\text{Z}}}_{i}-P^{\bar{\text{X}}\bar{\text{Z}}}_{i-1}$ is the probability that RUS terminates on exactly the $i$-th attempt. Note $P^{\bar{\text{X}}\bar{\text{Z}}}_{0}=0$. The value of $m$ allows one to extract the error on the $m$-th photon of an encoded qubit in the resource state. Odd numbers of $X$ ($Z$) photon error(s) would lead to a flip in $\overline{ZZ}$ ($\overline{XX}$).
    \item If RUS fusion ends up with only $\overline{ZZ}$, $m$ is sampled from the list $\mathcal{P}_{\bar{Z}}=\{P_i\}_{i\in [1,N]}$ where $P_i = P^{\bar{\text{Z}}}_{i}-P^{\bar{\text{Z}}}_{i-1}$. Note $P^{\bar{\text{Z}}}_{0}=0$. Errors in the photons involved in the $m$-th physical fusion would contribute to an error in $\overline{ZZ}$. 
    \item When a RUS fusion leads to $\overline{XX}$ only, $m$ is sampled from $\mathcal{P}_{\bar{X}}=\{P^{\bar{X}}_i\}_{i\in [1,N]}$. In such case, $\overline{XX}=(XX)^{\otimes m}$, and hence an odd number of errors in the first $m$ photons in each of the two encoded qubits would flip $\overline{XX}$.
\end{enumerate} 

\newpage
\section{Comparing theoretical thresholds with current state-of-the-arts}\label{supp:compare}

The threshold analysis presented in the main text is hardware-agnostic, as it focuses on error channels intrinsic to quantum emitters. Generally, any quantum emitter with access to spin and optical transitions (key prerequisites for the deterministic generation of entangled-photon cluster states) is inherently susceptible to spin dephasing, photon loss, and distinguishability errors. The fault-tolerant thresholds we have derived aim to provide a useful baseline, which helps consolidate experimental efforts towards improving specific areas crucial for practical implementation of the architecture. In this context, it is instructive to compare the current state-of-the-art figures of merits with the theoretical thresholds, as it highlights areas which require further advancements.

\begin{table*}[h]
\centering
    \begin{tabular}{|c|c|c|c|c|}
    \hline
     \textbf{Architecture} & \makecell{\textbf{End-to-end}\\ \textbf{efficiency}}{} & \makecell{\textbf{Emitter-emitter}\\ \textbf{indistinguishability}}{} & \multicolumn{2}{c|}{\textbf{Required $T_2$ (photon generation time $\tau$)}}\\
    \hline
    sFFCC, RUS ($N=10$) & $92\%$ & $96\%$ & \multicolumn{2}{c|}{$500\tau$} \\
    \hline\hline
    \multirow{2}{*}{\textbf{Emitter platform}} & \textbf{Outcoupling} & \textbf{Emitter-emitter} & \multicolumn{2}{c|}{\textbf{$T_2$ time}} \\ 
    \cline{4-5}
      & \textbf{efficiency}  & \textbf{indistinguishability}  & \textbf{Required} & \textbf{Achieved}\\
    \hline
    Neutral atom in optical cavity & $66\%$~\cite{Thomas2022} & $95.5\%$~\cite{vanLeent2022} & $37.5~$ms ($\tau=75~\mu$s~\cite{Thomas2024}) & $1.2$~ms~\cite{Thomas2022} \\
    Quantum dot & $57\%$~\cite{Tomm2021}, $71\%$~\cite{Ding2023} & $93\%$~\cite{Zhai2022} & $34.5~\mu$s ($\tau=68.9$~ns~\cite{Meng2024}) & $113~\mu$s~\cite{Zaporski2023} \\
    SiV in diamond cavity & $42\%$~\cite{Bhaskar2020} & $72\%$~\cite{Sipahigil2014} &  $0.071$~ms ($\tau=142$~ns~\cite{Bhaskar2020}) & $>1.5~$ms~\cite{Nguyen2019}, $13$~ms~\cite{Sukachev2017} \\
    \hline
    \end{tabular}
    \caption{\textbf{Comparison between experimental state-of-the-arts and thresholds required for sFFCC}. \textbf{Upper panel:} required specifications to reach fault-tolerance for sFFCC. It is important to note that these parameters are derived from the relevant thresholds assuming other errors are absent. \textbf{Lower panel:} Current state-of-the-arts. Outcoupling efficiency is considered to allow comparison between different material platforms. Emitter-emitter indistinguishability is defined by the measured raw Hong-Ou-Mandel visibility between photons from two emitters. }
    \label{tab:compare}
\end{table*}

The upper panel of Table~\ref{tab:compare} shows the required figures of merits derived from thresholds for the sFFCC architecture with RUS fusions and emitter re-initialization. The lower panel lists the experimental state-of-the-art parameters for three prominent material platforms, which have emitters with a built-in spin echo sequence. Note that the table is not meant to be exhaustive, and some numbers may not be directly comparable due to different experimental conditions. From the table, several conclusions could be drawn: (1) photon loss is the primary bottleneck for FBQC. The $92\%$ required end-to-end efficiency derived from photon loss threshold includes: the outcoupling efficiency of each resource state generator, fiber transmission, coupling efficiency to chip, and detection efficiency etc. While tremendous progress has been made in improving the efficiencies of integrated components~\cite{Alexander2024}, the outcoupling efficiency still requires significant advancements to bridge the gap. (2) Both single neutral atom and quantum dot systems have demonstrated close-to-threshold photon distinguishability between emitters. (3) Both quantum dot and SiV platforms may have reached the required tolerance for spin decoherence, while single neutral atom suffers from long generation time $\tau=75~\mu$s between two consecutively emitted photons owing to both qubit transfer and rotation pulses~\cite{Thomas2024}. However, for both quantum dot and SiV, additional decoherence processes like laser-induced or heating-induced dephasing, should be considered in future simulations to obtain a more realistic threshold. 

\end{document}